\documentclass[]{aa}
\usepackage{graphicx}
\usepackage{txfonts}
\usepackage{xcolor}
\usepackage{ulem}
\usepackage{url}
\usepackage{longtable}
\usepackage{multirow}
\usepackage{siunitx}
\usepackage{booktabs}
\usepackage{amsmath}
\usepackage[colorlinks=true,allcolors=blue]{hyperref}
\usepackage{orcidlink}

\newcommand{\mum}{$\mu$m\xspace}
\newcommand{\al}{$\alpha$\xspace}
\newcommand{\alz}{$\alpha_{Zemax}$\xspace}
\newcommand{\be}{$\beta$\xspace}
\newcommand{\lam}{$\lambda$\xspace}

\makeatletter
\renewcommand*\aa@pageof{, page \thepage{} of \pageref*{LastPage}}

\begin{document} 

\title{Geometric distortion and astrometric calibration of the JWST MIRI Medium Resolution Spectrometer}


   \author{Polychronis Patapis
          \inst{1}\orcidlink{0000-0001-8718-3732}
          \and
          Ioannis Argyriou\inst{2}\orcidlink{0000-0003-2820-1077}
          \and
          David~R.~Law\inst{3}\orcidlink{0000-0002-9402-186X}
          \and
        Adrian~M.~Glauser\inst{1}\orcidlink{0000-0001-9250-1547}
          \and 
          Alistair~Glasse\inst{4}\orcidlink{0000-0002-2041-2462}
          \and  Alvaro~Labiano\inst{5,6}\orcidlink{0000-0002-0690-8824}
          \and
          Javier~Álvarez-Márquez\inst{5}\orcidlink{0000-0002-7093-1877}
          \and
    Patrick~J.~Kavanagh\inst{7}\orcidlink{0000-0001-6872-2358}
          \and
          Danny~Gasman\inst{2}\orcidlink{0000-0002-1257-7742}
          \and
          Michael~Mueller\inst{8,9}
          \and
          Kirsten~Larson\inst{3}\orcidlink{0000-0003-3917-6460}
          \and
    Bart~Vandenbussche\inst{2}\orcidlink{0000-0002-1368-3109}
          \and      
          David~Lee\inst{4}
          \and
          Pamela Klaassen \inst{4}
          \and
          Pierre Guillard \inst{10,11}\orcidlink{0000-0002-2421-1350}
          \and
    Gillian~S.~Wright\inst{4}\orcidlink{0000-0001-7416-7936}       
          }

   \institute{Institute of Particle Physics and Astrophysics, ETH Zürich, Wolfgang-Pauli-Str 27, 8049 Zürich Switzerland
             \and
                Institute of Astronomy, KU Leuven,
              Celestijnenlaan 200D, 3001 Leuven, Belgium
              \and 
              Space Telescope Science Institute, 3700 San Martin Drive, Baltimore, MD, 21218, USA
              \and
              UK Astronomy Technology Centre, Royal Observatory, Blackford Hill Edinburgh, EH9 3HJ, Scotland, United Kingdom
              \and 
              Telespazio UK for the European Space Agency, ESAC, Camino Bajo del Castillo s/n, 28692 Villanueva de la Ca\~nada, Spain \label{tpz}
              \and
              Centro de Astrobiolog\'{\i}a (CAB), CSIC-INTA, Ctra. de Ajalvir km 4, Torrej\'on de Ardoz, E-28850, Madrid, Spain \label{cab}    \and
              Department of Experimental Physics, Maynooth University, Maynooth, Co. Kildare, Ireland
              \and
              SRON Netherlands Institute for Space Research, P.O. Box 800, 9700 AV Groningen, The Netherlands
              \and
              Sterrewacht Leiden, P.O. Box 9513, 2300 RA Leiden, The Netherlands
              \and 
              Sorbonne Universit\'{e}, CNRS, UMR 7095, Institut d'Astrophysique de Paris, 98bis bd Arago, 75014 Paris, France
              \and
              Institut Universitaire de France, Minist{\`e}re de l'Enseignement Sup{\'e}rieur et de la Recherche, 1 rue Descartes, 75231 Paris Cedex 05, France\\
        \email{polychronis.patapis@phys.ethz.ch}             }
   \date{Received July 3, 2023, Accepted xx}

 
  \abstract
   {The Medium-Resolution integral field Spectrometer (MRS) of MIRI on board JWST performs spectroscopy between 5 and 28~\mum, with a field of view varying from $\sim$13 to $\sim$56 arcsec square. The optics of the MRS introduce substantial distortion, and this needs to be rectified in order to reconstruct the observed astrophysical scene.}
   {We aim to use data from the JWST/MIRI commissioning and cycle 1 calibration phase, to derive the MRS geometric distortion and astrometric solution, a critical step in the calibration of MRS data. These solutions come in the form of transform matrices that map the detector pixels to spatial coordinates of a local MRS coordinate system called \al/\be, to the global JWST observatory coordinates V2/V3.}
   {For every MRS spectral band and each slice dispersed on the detector, the transform of detector pixels to \al/\be is fit by a two-dimensional polynomial, using a raster of point source observations. The dispersed trace of the point source on the detector is initially estimated by fitting a one-dimensional empirical function, and then iterating on the fist distortion solution using forward modelling of the point spread function model based on the \texttt{webbpsf} \texttt{python} package. A polynomial transform is used to map the coordinates from \al/\be to V2/V3.}
   {We calibrated the distortion of all 198 discrete slices of the MIRI/MRS IFUs, and derived an updated Field of View (FoV) for each MRS spectral band. The precision of the distortion solution is estimated to be better than one tenth of a spatial resolution element, with a root mean square (rms) of 10 milli-arcsecond (mas) at 5 \mum, to 23 mas at 27 \mum. Finally we find that the wheel positioning repeatability causes an additional astrometric error of rms 30 mas.}
   {We have demonstrated the MRS astrometric calibration strategy and analysis for all four integral field units, and all spectral bands of the MRS enabling the calibration of MRS spectra. This is a critical step in the data pipeline of every MRS observation, especially for science with spatially resolved objects. The distortion calibration was folded into the JWST pipeline in Calibration Reference Data System (CRDS) context jwst\_1094.pmap. The distortion calibration precision meets the pre-launch requirement, and the estimated total astrometric uncertainty is 50 mas. }

    \titlerunning{JWST/MIRI MRS astrometric calibration}

    \authorrunning{P. Patapis et al.} 

   \maketitle
%
\section{Introduction}
The Mid-Infrared Instrument \citep[MIRI,][]{Wright2015, Rieke2015a, Wright2023_MIRI} is one of the four science instruments on board the James Webb Space Telescope \citep[JWST,][]{Gardner2023}, and the only instrument operating in the mid-infrared. 
The MIRI Medium Resolution Spectrometer \citep[MRS,][]{Wells2015, Argyriou2023_MRS} is an integral field spectrometer (IFS) providing moderate resolution spectroscopy  \citep[R $\sim$ 4000 -- 1500,][]{Jones2023}, covering the wavelength range from 4.9 \mum\, -- 27.9 \mum \citep{Labiano2021, Argyriou2023_MRS}. In the near-infrared (0.6-5.3 \mum) the Near Infrared Spectrometer \citep[NIRSpec,][]{Boeker2023} also provides an IFS mode similar to the MRS.

\begin{figure*}[h]
    \centering
    \includegraphics[width=1.\hsize]{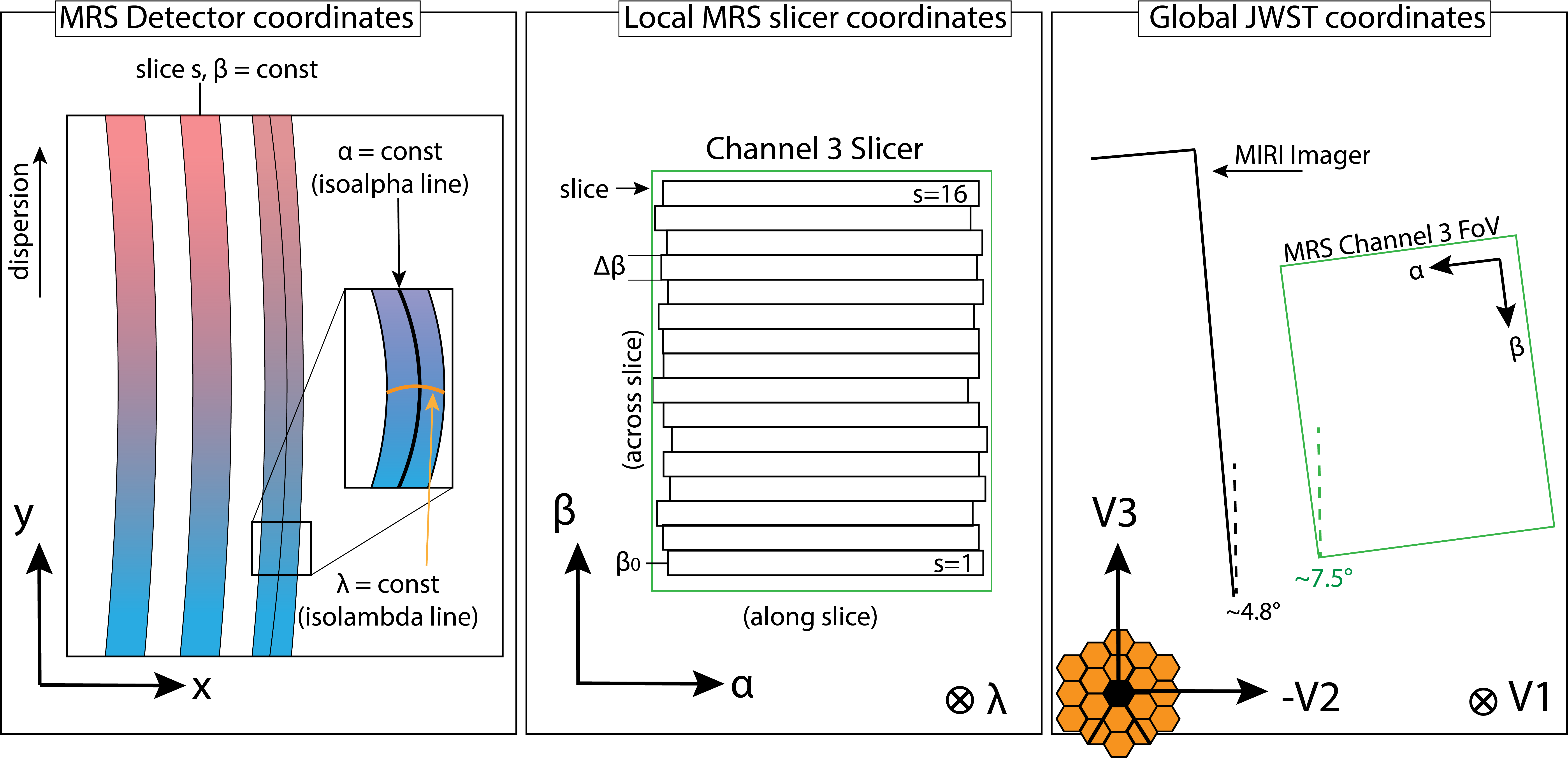} 
    \caption{Illustration of MRS coordinate systems. Left: the detector coordinates contain the 2D spectra projected by the IFU on the MIRI detectors as curved slices. Middle: the local MRS slicer coordinates are changing from band to band and represent the alignment of the slicer optic, with \al being the along-slice and \be the across-slice coordinate. Right: the JWST coordinate system V2/V3 is defined by the JWST Optical Telescope Element (OTE) and is shared among all instruments. Note that this panel is not to scale with the actual dimensions since the MRS FoV is very small.}
    \label{fig:MRScoords}
\end{figure*}

The advantage of an IFS is that it retains information for both spatial coordinates of the astronomical scene being observed using image slicers \citep{Allington-Smith2006}. The MRS employs one image slicer for each spectral channel. Each image slicer splits one of the spatial dimensions into distinct slices before dispersing the light. The number of slices depends on the physical geometry of the image slicer, optimised for the spectral and spatial resolution element of each MRS channel. The data that are imaged on the MRS detectors are akin to having multiple single slit spectra, one for each slice in a spectral channel. In order to reconstruct the scene observed on the sky one must map each pixel of the detector to a position in right ascension (RA), declination (DEC) and wavelength, accounting for the optical distortion that is typically present in IFS. 

In this work we present the derivation of the geometric distortion and astrometric calibration of the MRS, an essential product of the JWST data processing pipeline. The MRS is spatially and spectrally under-sampled at all wavelengths by design, and requires at least two dithered exposures to be Nyquist sampled. The sampling artefacts and cube reconstruction algorithm are detailed in \cite{Law2023}. The optical distortion itself needs to be corrected to subpixel precision, since even small residuals can introduce systematic errors in the science products, given the under-sampled PSF core. A well calibrated distortion over the full FoV and spectral range is required to maximise the optical quality of the instrument, and restore the diffraction limited performance provided by JWST that is distorted by the optics of the MRS. An accurate plate scale enables a consistent spectro-photometric calibration over the full FoV and reduces errors in the surface brightness estimation for extended sources \citep{Gordon2022}. 

In Sect.~\ref{sec:mrs_intro} we introduce the relevant coordinate frames and transforms used in the JWST calibration pipeline, and discuss the origin of the optical distortion in the MRS. In Sect.~\ref{sec:data} we describe the different data sources used to derive the distortion calibration, and the methods used to estimate the point source traces on the detector. In Sect.~\ref{sec:flight} we detail the analysis steps that were used to derive the geometric and astrometric distortion transforms. Finally, in Sect.~\ref{sec:discussion} we discuss the precision of the distortion solution achieved through this work, the astrometric stability due to the repeatability of the grating wheel, and lessons learned for the astrometric calibration of an IFU instrument for application in future projects.

\section{MIRI Medium Resolution Spectrometer} \label{sec:mrs_intro}

\subsection{Brief Optical Description of the MRS}
 The light enters the MRS from the JWST focal plane via a pickoff mirror and is relayed to a series of three dichroic filters that split the light into the four MRS spectral channels (1 -- 4). The dichroic filters are always configured in such a way that the same sub-band SHORT/MEDIUM/LONG or A/B/C is observed at one time\footnote{For sub-band B, the MRS will simultaneously observe spectral bands 1B, 2B, 3B, and 4B.}. The dichroic-filtered light is fed to the input of the four MRS integral field units (IFUs), while two blocking filters and corresponding light traps prevent unwanted stray-light from entering the spectrometer. 
 
When entering the IFU of one of the four MRS spectral channels the input focal plane is anamorphically magnified and reimaged onto the image slicer of that channel. The image slicer consists of thin mirrors, diamond turned onto a common substrate in a pyramid like manner (see Fig. 6 in \cite{Wells2015}). Each channel has a different number of slicing mirrors, with 21, 17, 16 and 12 for channels 1, 2, 3 and 4 respectively. Each slice is separated spatially by the angle of the slicer mirrors towards a re-imaging mirror with a pupil mask placed in the intermediate pupil plane to control the stray-light. The re-imaging mirrors create an image of each slice on a slitlet mask that defines the output of the spectral channel IFU. The beams from the slits are then collimated by a mirror and diffracted by the diffraction grating that is located on the reverse side of the same wheel as the dichroic filter that transmitted the wavelengths for the specific spectral channel. The MRS incorporates two wheels, denoted as the dichroic grating wheel assembly A (DGA-A) and dichroic grating wheel assembly B (DGA-B). The first diffraction order of the grating-diffracted beam is imaged onto the detector. The MRS camera optics combine the beams of two channels, 1 and 2, and 3 and 4, and focus the light onto the short wavelength (MIRIFUSHORT) and the long wavelength (MIRIFULONG) Si:As IBC detectors \citep{Rieke2015b,argyriou2020SPIE} respectively.

\subsection{MRS Coordinate Systems}

There are three coordinate systems that are relevant for this work and the MRS, illustrated in Fig.~\ref{fig:MRScoords}. These are (i) the MRS detector coordinates, (ii) the local MRS coordinates, and (iii) the JWST telescope coordinates which are connected to the right ascension (RA) and declination (DEC) of an astronomical object in the sky.

(i) The detector coordinates are defined by the pixels of the detector arrays that have a dimension of (1032, 1024). The horizontal axis of the detector is roughly aligned with the MRS IFU image slicer along-slice (denoted as \al) spatial coordinate, and the vertical axis is roughly aligned with the dispersion direction, shown on the left panel of Fig.~\ref{fig:MRScoords}. We note that we assume zero-indexed arrays and therefore define the center of the lower left pixel as (x, y) = (0, 0). 

(ii) The local MRS coordinates are aligned to the image slicer along- and across-slice direction and are denoted as local due to the fact that they are unique to each MRS sub-band. Due to small alignment differences of the slicer optic and the dichroic filters, the slicer location of each MRS sub-band projected on the sky are not perfectly concentric, yielding small boresight and rotation offsets. The along-slice coordinate is defined as \al and the across-slice coordinate as \be, and both have units of arcseconds. The coordinate \be is often used interchangeably with the term "slice", referring to an individual sliced image created by the slicer and dispersed onto the detector. While not having a physical connection to the slicer, a third coordinate referring to the wavelength and denoted as \lam, is often used in conjunction with \al/\be  to complete the three dimensional coordinate system of the MRS IFU. In the context of the overall MRS instrument calibration, like optical distortion, wavelength calibration, fringing, straylight and PSF \citep{Wells2015, Argyriou2020a, Labiano2021}, and together with the detector coordinates, the vector (\al, \be, \lam) operates as a self-contained coordinate system as it offers an intuitive view of the optics.

(iii) The JWST Observatory coordinate system is defined and shared among all instruments. It is defined by two orthogonal coordinates on the JWST primary aperture plane called V3, that points towards the secondary Mirror Support Structure, and V2 orthogonal to V3 \footnote{\url{https://jwst-docs.stsci.edu/jwst-observatory-characteristics/jwst-observatory-coordinate-system-and-field-of-regard}}, as shown in the right panel of Fig.~\ref{fig:MRScoords}. A third coordinate V1 that is the telescope symmetry axis completes the coordinate system. These coordinates enable the observatory to slew towards a target on sky and align the target with the instrument selected for the observation. Additionally, it enables dithering strategies and pointing offsets that are essential to some observing modes like coronagraphy and IFU spectroscopy. The (V2, V3) coordinates have units of arcseconds. The JWST coordinate system connectes to RA, DEC through the V3 position angle (PA), that measures the rotation of the observatory with respect to north when projected on sky. All this information is contained in the JWST Science Instrument Aperture File (SIAF) on board the observatory, with V2/V3 also often referred to as SIAF coordinates. The MRS FoV is located on the far right of the JWST FoV at a distance of 20 arcseconds from the top right corner of the MIRI Imager, as seen in the right panel of Fig.~\ref{fig:MRScoords}. The FoV of MIRI are rotated with respect to V3 by an angle of $\sim 4.8^{\circ}$ for the Imager, and $\sim 7.6^{\circ}$ to $\sim 8.8^{\circ}$ for the MRS, as shown in Fig.~\ref{fig:MRScoords}.

\subsection{Geometric Distortion}

Ideally the optics would disperse and image the spectra of each slice orthogonal to the detector, and similar to most conventional spectrographs, this is not the case for the MRS. The anamorphic optics, slicer and camera optics introduce significant optical distortion to the imaged field, resulting in the spectra being projected onto the detector as curved lines. This curvature is illustrated in the left panel of Fig.~\ref{fig:MRScoords}, and is typical for slit spectrometers, usually referred to as keystone and smile distortion \citep{Yokoya2010}. The distortion in the along-slice spatial direction is more subtle than just a curved spectrum on the detector; the plate scale (detector pixel subtended angle on the sky), is changing non-linearly as a function of position in the field and wavelength, differently for each slice. This intra-slice distortion appears in both the dispersion and the spatial direction, and the rectification of these coordinates, from detector pixels to MRS local coordinates (\al, \be, \lam), is a crucial step in the JWST calibration pipeline in order to reconstruct the observed astronomical scene and conserve the optical quality of the MRS. 

We define two terms. First, lines on the detector that trace a constant value of \al are denoted as iso-\al lines. Second, lines that trace constant wavelength (\lam) are denoted as iso-\lam lines (the wavelength distortion calibration based on ground test data is described in detail by \cite{Labiano2021}). The transformation of pixel coordinates to local MRS coordinates is referred to as a "detector-to-cube" transformation, and is described by a set of polynomial transforms for each sub-band (1A to 4C). Each transform maps the coordinates (x, y) to \al, \lam as:

\begin{equation}\label{eq:alpha_polynomial}
    \alpha_s(x, y) = \sum^{N, N}_{i, j} K_{\alpha, s}(i, j) (x-x_s)^j y^i ,
\end{equation}
\begin{equation}
   \lambda_s(x, y) = \sum^{N, N}_{i, j} K_{\lambda, s}(i, j) (x-x_s)^j y^i , 
\end{equation}

with N being the order of the two-dimensional polynomial transform, $K_{\alpha, s}$ and $K_{\lambda, s}$ the polynomial coefficient matrices and $x_s$ a reference pixel in the middle of each slice. The \be coordinate is discrete and only depends on the slice number since it is collapsed when dispersing the light on the detector. The value of \be is given by Eq.~\ref{eq:beta}.

\begin{equation}\label{eq:beta}
    \beta(s) = \beta_0 +(s-1)\Delta\beta,
\end{equation}

where $\beta_0$ is the \be coordinate of the center of slice 1 and $\Delta\beta$ the slice width of each channel in arcseconds. The specific values for each of the MRS spectral channels are tabulated in Table~\ref{tab:betaFoV}, taken from \cite{Wells2015}. We define $s=1$ at the center of the first slice, with the edges of the slice being $s \pm 0.5$. 

\begin{table}[h]
    \centering
      \caption{MRS across slice parameters. Values are taken from \cite{Wells2015}.}
    \begin{tabular}{c|c|c| c}
        Channel & \# slices & $\Delta \beta$ [arcsec] & $\beta_0$ [arcsec] \\ \hline
        1 & 21 & 0.177 & -1.77 \\
        2 & 17 & 0.280 & -2.24 \\
        3 & 16 & 0.390 & -2.92 \\
        4 & 12 & 0.656 & -3.61 \\
    \bottomrule
    \end{tabular}
  
    \label{tab:betaFoV}
\end{table}

There are two transformations to be made. The first is from detector pixels (x, y, $s$) to (\al, \be, \lam) to account for the MRS specific distortion of each sub-band, and in a second step we transform the local MRS coordinates (\al, \be, \lam) to the JWST global coordinates (V2/V3). This second transform accounts for distortion introduced by the JWST Optical Telescope Element (OTE) which all instruments are subject to, as well as placing the FoV of the MRS (\al, \be), that is slightly rotated, onto V2/V3. This is shown in the right panel of Fig.~\ref{fig:MRScoords}. The coordinate transform from (\al, \be) to V2/V3 is given by a second order polynomial:

\begin{equation} \label{eq:v2_trans}
    V2_{Ch}(\alpha, \beta) = \sum_{i,j=0, 0}^{2, 2} T_{Ch, V2}(i, j) \alpha^j \beta^i
\end{equation}
\begin{equation}\label{eq:v3_trans}
    V3_{Ch}(\alpha, \beta) = \sum_{i,j=0, 0}^{2, 2} T_{Ch, V3}(i, j) \alpha^j \beta^i,
\end{equation}

where $T_{Ch, V2},\, T_{Ch, V3}$ are the polynomial coefficient matrices for V2 and V3 respectively. 

\section{Data and Methods} \label{sec:data}

\subsection{Commissioning and Cycle 1 Calibration}

During JWST commissioning dedicated observations listed in Table~\ref{tab:com_data} were executed, in order to validate and update the distortion of MIRI. For the MRS the goal was to check the geometric distortion, derive the field transform from local MRS coordinates to the JWST SIAF, update the boresight offsets with respect to the MIRI Imager, and test the MRS dither patterns. The first program that was used is JWST Program Identifier (PID) 1012, consisting of exposures of the MRS internal calibration lamp, which fully illuminates the detector slices and was used to derive the detector slice mask.

To help derive the astrometric calibration, the strategy was to observe bright stars with the MRS, including simultaneous MIRI Imaging and parallel FGS observations. Both FGS and the MIRI Imager, with their large FoV, contained many stars with very precise astrometric information from GAIA \citep{gaia1,gaia2}. PID 1049, and 1050 were part of the PSF measurement and provided very bright sources for channels 3/4 and channels 1/2 respectively. For PID 1049 a red planetary nebula (SMP LMC-58) was observed in the dither patterns optimized for the long channels (3, 4) of the MRS. In PID 1050 a photometric standard A-star (HD 163466) was observed in the point source optimised dither pattern of the channel 1, as well in an extended source dither pattern and at the instrument boresight.

\begin{figure}[p]
    \centering
    \includegraphics[width=0.77\columnwidth]{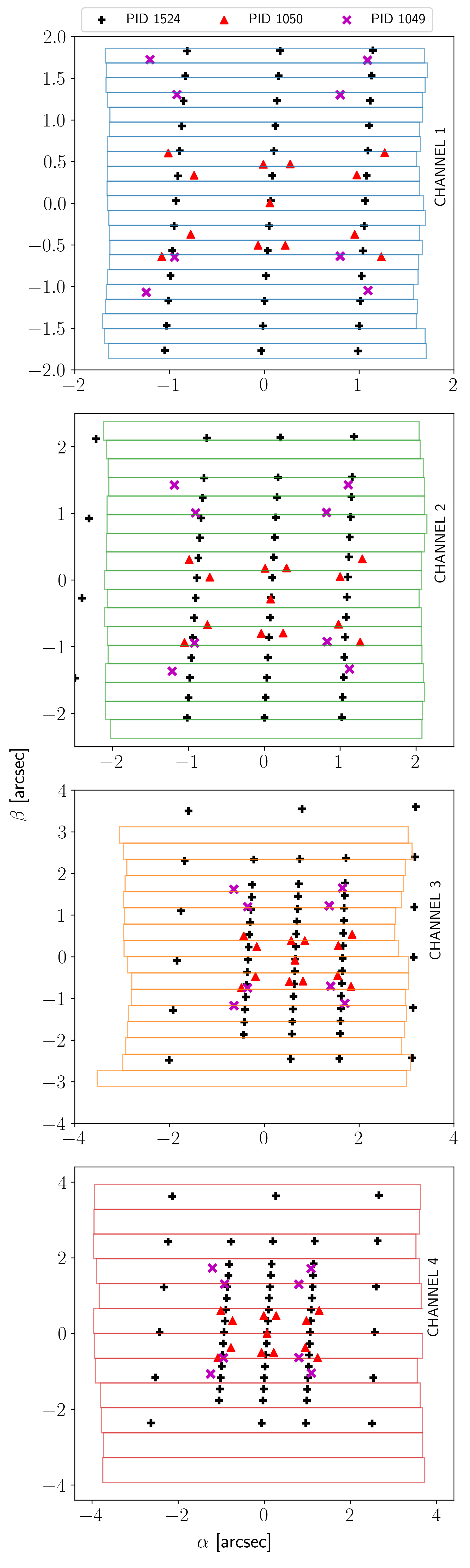}
    \caption{Dither points from commissioning and cycle 1 calibration, with each panel illustrating the outline of each channel's slicer. The black crosses correspond to the dither pattern of 10~Lac executed during PID 1524 (observation 16), optimised to characterise the intra-slice distortion. The red points correspond to the point and extended dither pattern optimised for channel 1 (PID 1050, observation 9), and the violet points correspond to the point source dither pattern of channel 3 (PID 1049, observation 2).}
    \label{fig:cal24_raster}
\end{figure}

As described in Section~\ref{sec:flight}, additional observations were needed in order to improve the distortion solution of the MRS. A Cycle 1 calibration program (PID 1524, observation 16) was designed and executed as one of the first calibration programs post-commissioning, with the goal to characterise the MRS distortion in detail. A custom dither pattern of a bright O-star, 10~Lac, was uploaded to the observatory, which would place the point source to three or more positions in the along-slice direction \al for most slices of each channel. This enables to fit the plate scale for each slice with the required second order polynomial \citep{Glauser2010}. The custom dither pattern that includes 57 points is shown in Fig.~\ref{fig:cal24_raster}, which shows its overlap with the slicer of each channel. 10~Lac was chosen as the target since it was bright enough to be efficiently observed with 20 frames per integration, provided sufficient S/N up to channel 4C, and had emission lines that could be used to calibrate the wavelength distortion of the MRS. Finally, subsequent observations of 10~Lac (PID 1524, observation 17) and observations of standard A and G stars through Cycle 1 photometric calibration programs (1536, 1538) that included target acquisition (TA) enabled the astrometric precision monitoring of the MRS.

\begin{table}[t]
    \centering
    \caption{Commissioning and Cycle 1 calibration programs and stars used for the in-flight calibration.}
    \begin{tabular}{l l l l}
        Program ID & Name & Type  & K mag\\ \hline
        1024 & 2M J05220207-6930388 & LMC star & 10.0\\
        1049 & SMP LMC-58 & PN & 14.5 \\
        1050 & HD 163466 & A2   & 5.1 \\
        1050 & HD 159222 & G1  & 4.9 \\
        1524 & 10~Lac & O9  & 5.5 \\
        1536 & HD2811 & A3V & 7.0 \\
        1536 & del UMi & A1  & 4.2 \\
        1538 & 16 Cyg B & G3V  & 4.6 \\
        1538 & HD 167060 & G3V  & 7.4 \\
        1538 & HD 37962 & G2V & 6.2 \\
        \hline 
        1012 & \multicolumn{2}{c}{Internal calibration lamp} &  - \\
    \bottomrule
    \end{tabular}
    
    \label{tab:com_data}
\end{table}
 
The standard JWST MIRI MRS pipeline\footnote{jwst pipeline version 1.9, Calibration Reference Data System (CRDS) version: 11.16.3, CRDS context: jwst\_0932.pmap}\citep{Labiano2016, bushouse_howard_2022_6984366} was used to process the raw data files from the observations. First the \texttt{Detector1Pipeline} was ran to obtain rate files , i.e. slope images \citep{morrison23}. From the \texttt{Spec2Pipeline} that deals with the calibration of spectroscopic modes of JWST (for the MRS see \cite{Argyriou2023_MRS}), we used the scattered light and detector fringing effects. The distortion and astrometric calibration folds into the first step of \texttt{Spec2Pipeline}, that is the \texttt{assign\_wcs}. Dedicated background observations for each observation were subtracted on the detector level as slope images. 

\subsection{Detector-based point source tracing}\label{sec:det_fit}

\begin{figure*}
    \centering
    \includegraphics[width=\textwidth]{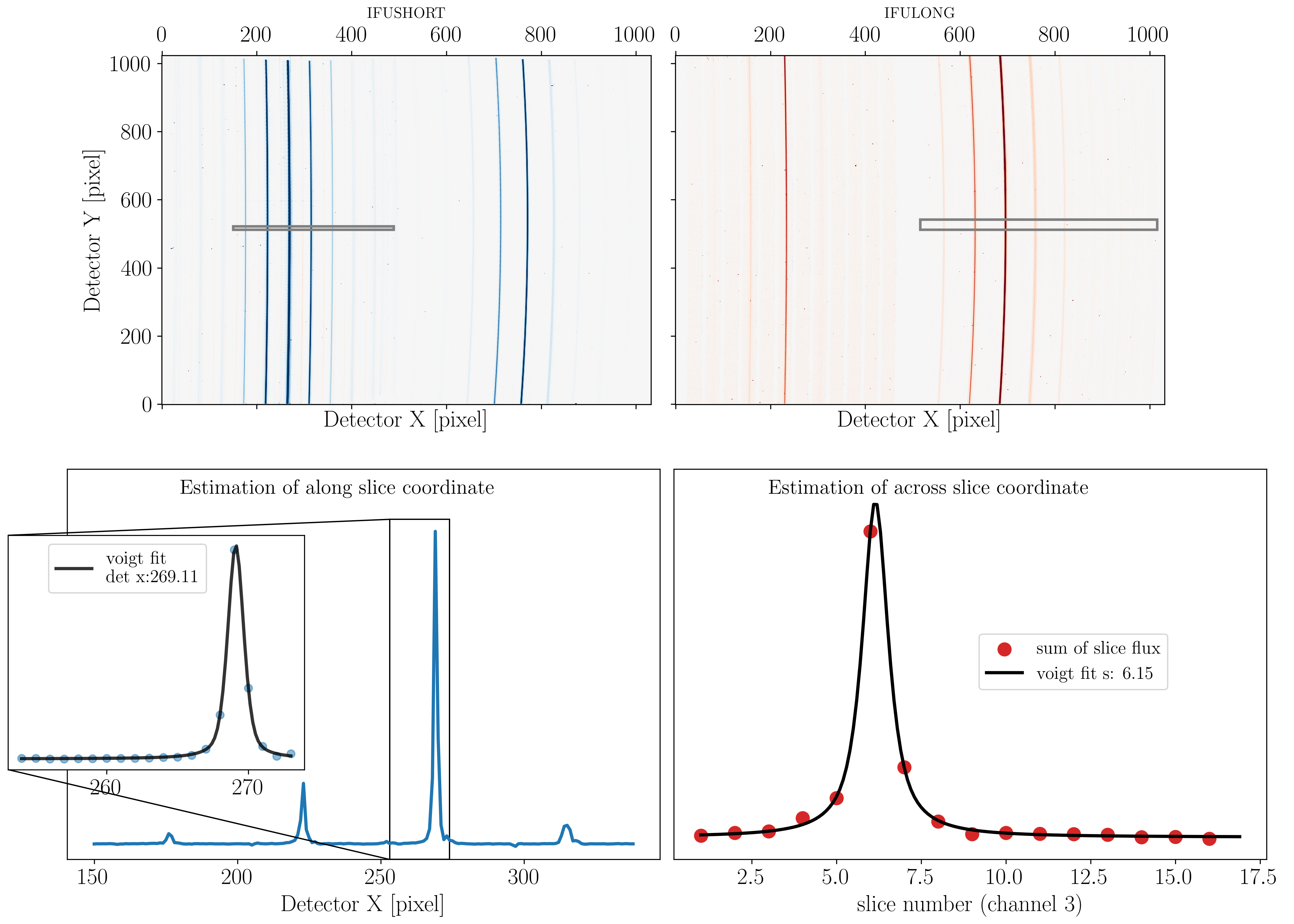}
    \caption{Detector tracing for MRS point sources. Top row: Detector images of bright star 10 Lac from PID 1524 showing left the IFUSHORT detector (Ch.~1 and Ch.~2.), and right IFULONG (Ch.~3 and Ch.~4.). Bottom left: Single row of detector showing the PSF projected on it, and fit with pseudo-Voigt function (inset) to estimate the centre in the detector x (corresponding to \al) coordinate. The centre of the trace for the brightest slice (core of the PSF) coincides with the centre of the PSF. Bottom right: the collapsed signal over a given number of detector rows in each slice is fitted with a pseudo-Voigt profle to estimate the across slice coordinate s (equivalent to \be).}
    \label{fig:det_tracing}
\end{figure*}

Here, we briefly present the methods used to trace a point source on the detector of the MRS, also shown in Fig.~\ref{fig:det_tracing}. These form the basis of the whole subsequent distortion calibration. There are two quantities that are extracted from the detector: (i) the across-slice position \be of the point source, and (ii) the along-slice, iso-\al trace as a function of detector position (x, y).

(i) In the across-slice direction the point source is sampled by the IFU image slicer mirror slices and the goal of the detector-fitting is to estimate the sub-slice position (or slicer coordinate \be) of the source. The signal in each slice on the detector is summed, to effectively integrate over the along-slice direction \al, and fitted against the slice index $s$ with a pseudo-Voigt profile\footnote{The pseudo-Voigt function, a linear combination of a Gaussian and Lorentzian}. The validity of this method was tested by taking a theoretical PSF model, sampling it by the number and width of the slices for each MRS band, and comparing the estimated parameter with the input coordinate. The error of the fit was in the simulated tests was in the order of a few percent of a slice width. In the bottom right panel of Fig.~\ref{fig:det_tracing}, an example of the slicer coordinate estimation in Ch.~3 using the bright star from PID 1524 is shown. 

(ii) The iso-\al can be traced in each slice in two different ways. First, using an empirical PSF profile fit directly on the detector signal. This can be done in a row-by-row manner, or along  an iso-\lam, with the pixels belonging to a given wavelength bin of a slice identified and fitted by the profile. Both a Gaussian and a pseudo-Voigt function were tested with the Voigt function resulting in a better fit (shown in \cite{Labiano2021},  Fig.~\ref{fig:det_tracing}). The pseudo-Voigt function is motivated by the fact that the light scattered within the MIRI detectors \citep{Gaspar2021} produces elongated wings for the MRS PSF \citep{Argyriou2023_MRS,Patapis2023_PSF}, and a similar behaviour is also seen in the MIRI Imager \citep{Dicken2023_MIRI}. For slices that contain a significant fraction of the PSF core the iso-\al traces are well fitted this way, since the PSF is symmetric, the centre of the empirical profile coincides with the centre of the PSF. 

\begin{figure*}[h]
    \centering
    \includegraphics[width=0.8\textwidth]{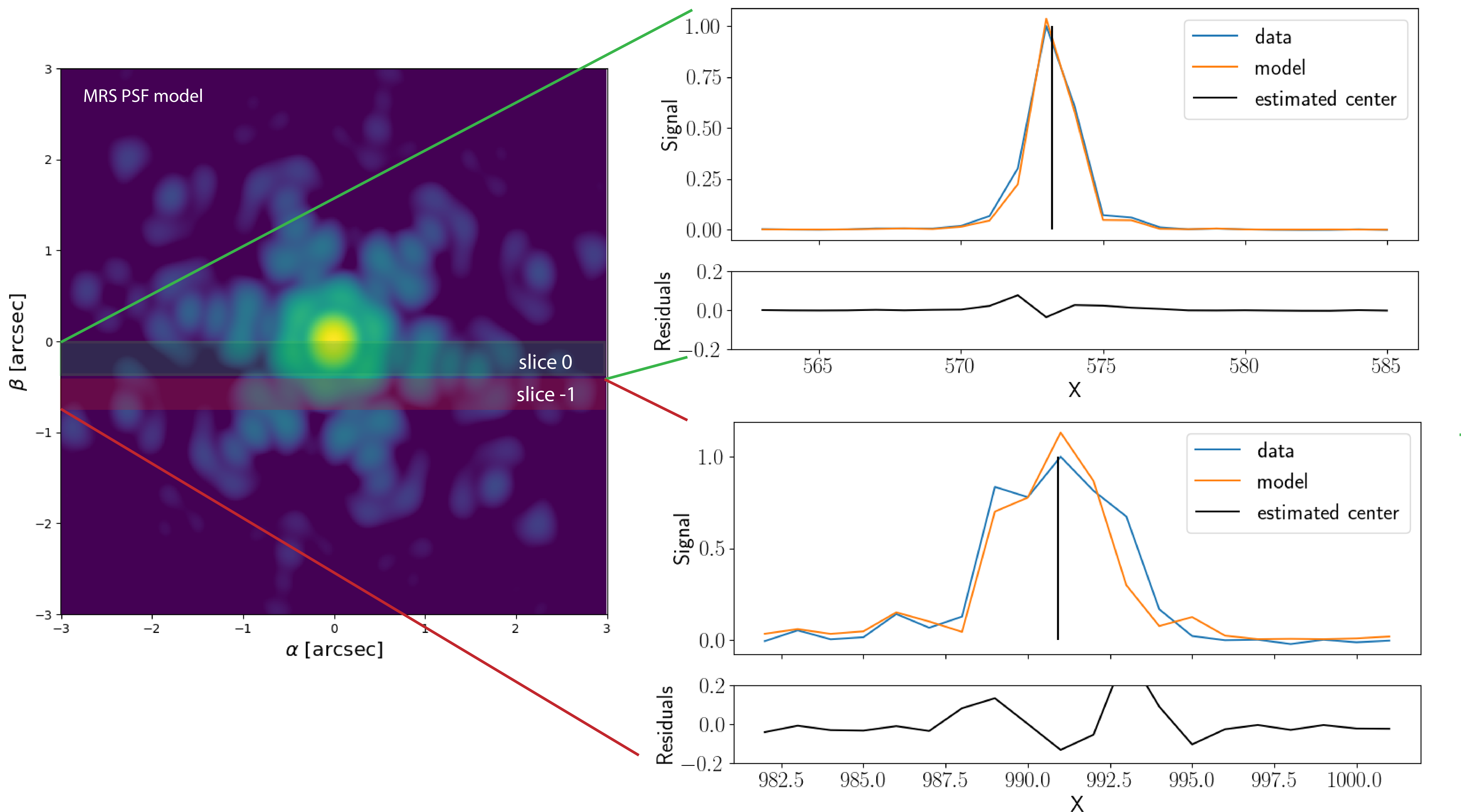}
    \caption{Forward modeling of the MRS webbPSF used for the estimation of iso-\al trace. The detector trace is determined by collapsing the corresponding slice in the webbPSF model along \be and projecting the \al profile of the PSF to the detector using the estimated distortion solution. The centre of the webbPSF model is optimised to match the data on the detector and estimate the centre of the given point source exposure. For well centred slices this works well, while for slices further from the PSF centre, the discrepancy between model and data is larger, however the source centre estimation is not affected significantly.}
    \label{fig:webbPSF_projection}
\end{figure*}

For slices illuminated by the PSF wings and not containing a significant part of the PSF core (mainly in channels 3, 4) we estimate the iso-\al traces on the detector using forward modeling of the theoretical MRS PSF as shown in Fig.~\ref{fig:webbPSF_projection}. The MRS PSF was modeled using the \texttt{python} package \texttt{webbpsf} \citep{Perrin2016}, and was broadened by convolution with a Gaussian kernel to match the optical quality observed in flight. The detailed description of this model and the commissioning PSF analysis for the MRS is described in \cite{Patapis2023_PSF}. The PSF model is then interpolated onto the detector using the distortion model, and iterating with each update of the distortion. In Fig.~\ref{fig:webbPSF_projection} we show that the PSF model, and distortion solution from ground are sufficient to reproduce the profile of a point source on the detector within 5\% for slices close to the PSF core (top right panel in Fig.~\ref{fig:webbPSF_projection}), and worsening to residual error of $\sim$20\% for slices further away (bottom right panel in Fig.~\ref{fig:webbPSF_projection}). Even with higher residuals, the error on the centre is in the order o 10\% of a pixel. This validity of the fitting was tested for multiple locations in the FoV, different rows on the detector, and for all bands. 

\section{In-flight astrometric and distortion calibration} \label{sec:flight}

\subsection{Systematic errors from ground calibration}

A preliminary calibration of the MRS geometric distortion was performed using the ground calibration data \citep{Glauser2010, Wells2015}, based on the optical design model of the MRS (modeled using the ray tracing software Zemax OpticStudio). That original analysis was performed using reconstructed cubes - median-collapsed in wavelength - while applying slice to slice corrections. The solution from the tweaked Zemax OpticStudio model was precise to approximately 0.5-1 resolution element depending on the band, however, the following issues observed in ground and flight data motivated a new derivation of the distortion solution for the MRS. 

First, the initial calibration used the cubes median-collapsed in wavelength to correct the MRS optical design model. This collapsing in wavelength left a significant systematic error in \al as a function of wavelength, of the order of a pixel (see Sect. 3.2.3 in \citet{phdthesisYannis}). The systematic error was partially corrected by re-fitting the ground data on the detector level which mitigated the issue (see Appendix~\ref{ap:ground_calibration}). However, when tested on flight data, at the top of the detector and for all available points a similar systematic was present, underestimating the iso-\al trace compared to the middle and bottom of the detector by half a pixel. With comparisons between bands pointing towards a global distortion systematic error, it potentially stems from issues in the optics of the test illumination source setup used in ground testing.


Second, by comparing the commanded offsets in V2/V3 coordinates with the ones estimated from fitting the point source centre on the detector, we found misalignment between slices in some bands, while in other bands (especially those in channel 3) there was evident magnification errors within the slices. All these systematic errors are probably tied to discrepancies in the estimated coordinate transforms using the ground test data, given the low S/N, non diffraction-limited and asymmetric PSF and distortion of optical test setup, and lack of point source data for more recent observatory level campaigns. These errors are of the order of 1 pixel. 

In order to minimize these systematic residuals in the alpha distortion calibration, and to derive the astrometric calibration that maps the local MRS coordinates to the V2/V3 coordinate frame, we use the PID 1524 cycle 1 calibration observations of the O-star 10~Lac shown in Fig.~\ref{fig:cal24_raster}. The updated distortion transforms are then tested on the commissioning observations in order to estimate the residual error in the distortion, as well as the estimated absolute astrometric error.


\subsection{Correction of telescope attitude matrix}
Although target acquisition can reliably place a star in the MIRI MRS to an accuracy of about 30~mas\footnote{\url{https://jwst-docs.stsci.edu/jwst-mid-infrared-instrument/miri-operations/miri-target-acquisition/miri-mrs-target-acquisition}}, the absolute position information of the world coordinate solution (WCS) association with any given observation is typically only accurate to about 300~mas due to a combination of errors in the guide star catalog and uncertainty in the spacecraft roll angle.  For the purposes of calibrating the absolute location of the MRS within the telescope focal plane we therefore obtained simultaneous FGS and MIRI imaging data in parallel with the dithered MRS observations.

Using 25 bright stars in the FGS imaging field with Gaia Data Release 3 \citep{gaia2} astrometry we re-derived the telescope attitude matrix for each of our exposures, improving the absolute positional accuracy of the WCS to 30~mas.  The FGS and MIRI imaging data likewise were used to confirm that the relative accuracy of the commanded dither offsets was good to 10~mas or better, which was deemed more than sufficient for our analysis.

\subsection{Detector Slice Mask}\label{sec:slice_mask}

We begin the analysis by deriving the detector slice mask that maps the projected slits on the MRS detectors. For a spatially homogeneous extended source illumination of the MRS, the signal within a slice is flat across the MRS FoV, dropping off rapidly at the edges. This allows for the derivation of the FoV limits along these slices (x-direction) for each detector row. We do so using solely the signal values on the detector, no other calibration information is required.

To derive the FoV limits on the detector plane for each slice, one has to measure the drop in throughput at the slice edges. The deviation of the overall throughput in each slice from an ideal boxcar function, however, makes this a non-trivial challenge. The derivation of an accurate slice mask is thus limited by the accurate definition of the throughput at the slice edges.

\begin{figure}[h]
\centering
\includegraphics[width=0.5\textwidth]{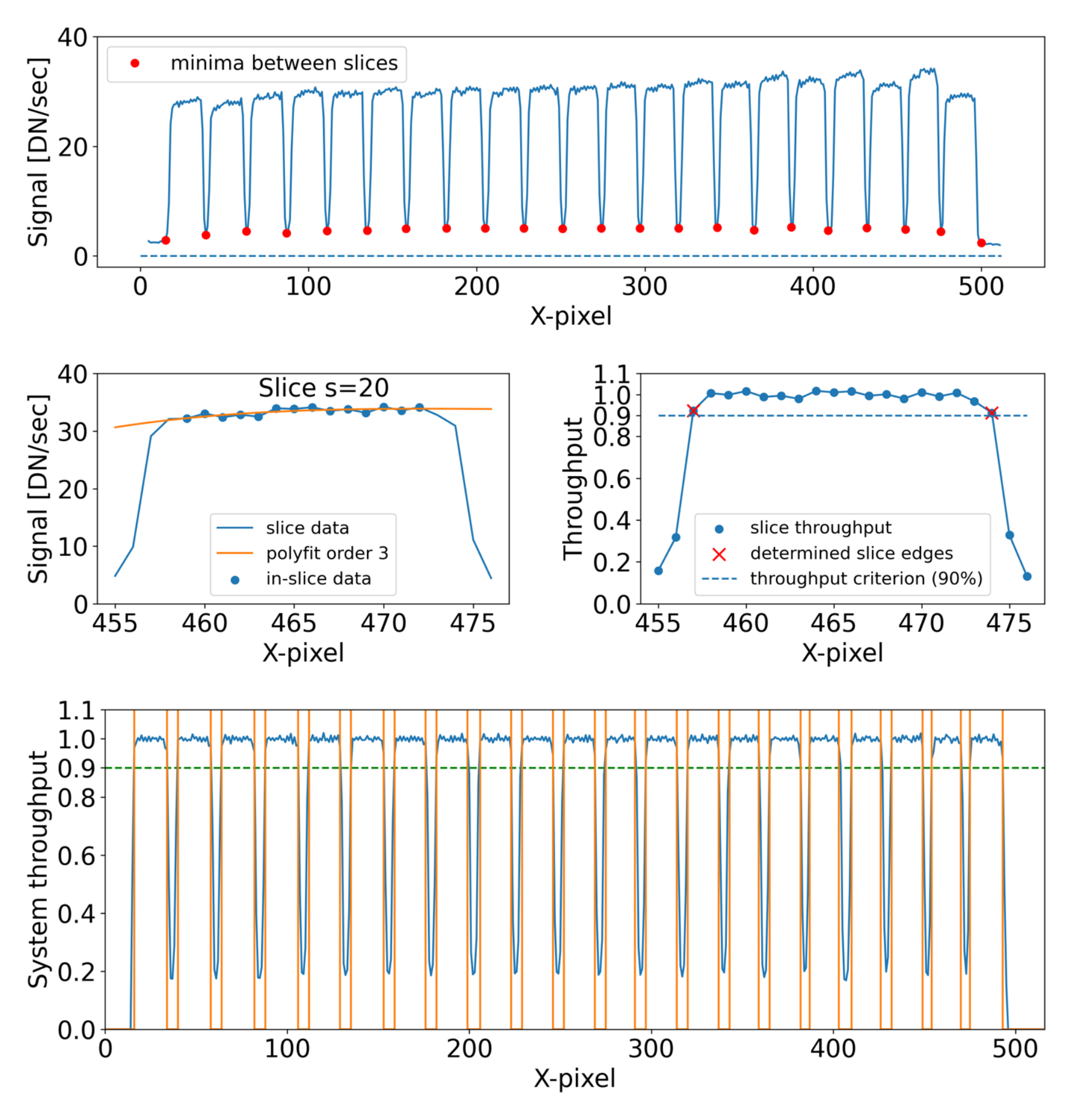}
\caption{Slice masking process based on internal calibration source exposures. Top: Signal of single detector row of calibration source, and finding the minima of the extended illumination correspond to the slice gaps. Middle: normalising the illumination in order to calculate the transmission factors. Bottom: with the normalised slices the pixel locations of various transmission levels for each slice and each row are calculated.}
\label{fig:slice_masking_process}
\end{figure}

The algorithm to derive a slice mask performs the following three steps for all slices in a MRS channel for each detector row. Firstly, the minima (troughs) in the signal separating the slices are identified as shown in the top plot of Fig.~\ref{fig:slice_masking_process}. These minima bound the slices in each channel. Secondly, the throughput of each slice is determined by first fitting a 3rd order polynomial to the signal values of the pixels that are well within the slice (5 pixels from the slice edges). This is shown in the middle right plot of Fig.~\ref{fig:slice_masking_process}. The signal between two minima surrounding the slice is then divided by the fitted 3rd order polynomial, this yields an estimation of the slice throughput as shown middle left plot of Fig.~\ref{fig:slice_masking_process}. Thirdly, a throughput criterion is used to define the left-most and right-most pixel that contributes to a slice (edges of the MRS FOV). We show the result for one detector row in channel~1, in the bottom plot of Fig.~\ref{fig:slice_masking_process}.

Iterating the above steps for each detector row produces a 2D slice mask. In fact, we define not one, but nine slice masks using this method. These nine slice masks correspond to different slice throughput levels, from 10\% throughput to 90\% throughput. Using a slice mask with 90\% throughput is a conservative approach that yields a conservative field of view. It is also possible to use a slice mask with a 50\% throughput. This will yield a larger MRS FoV, however, the signal at the FoV edges will be much lower. 
The JWST MIRI/MRS pipeline uses an 80\% transmission cut-off by default. 

\subsection{Field of view definition}\label{sec:flightFoV}

As seen in Fig.~\ref{fig:cal24_raster} not every slice of the dither pattern in PID 1524 (observation 16) has point sources centred on it.  We identified which exposures had signal in a given slice by iterating through all exposures, and finding the three along \al that had the maximum signal in the slice. These exposures were then selected to calculate the V2/V3 to \al/\be transform. We fitted their slicer coordinate on the detector using the pseudo-Voigt profile fitting shown in Fig.~\ref{fig:det_tracing}. Then, using the matrix form of a coordinate rotation Eq.~\ref{eq:field_transform1} and Eq.~\ref{eq:field_transform2}, the FoV was defined based on the V2/V3 coordinates reported in the metadata. 

\begin{equation} \label{eq:field_transform1}
    \begin{bmatrix}
    \alpha \\
    \beta
    \end{bmatrix}  
    =\begin{bmatrix}
     \alpha \\
     \beta_0 + (s-1)*\Delta \beta
    \end{bmatrix}\\
    =\begin{bmatrix}
     cos(\theta) & sin(\theta) \\
     -sin(\theta) & cos(\theta)
    \end{bmatrix}
    \begin{bmatrix}
    V2-V2_{0}\\
    V3-V3_{0}
    \end{bmatrix}
\end{equation}

\begin{equation} \label{eq:field_transform2}
    \beta_0 + (s-1)*\Delta \beta = -sin(\theta)(V2-V2_{0}) + cos(\theta)(V3-V3_{0})
\end{equation}

This assumes that the V2/V3 field is not significantly distorted over the size of the FoV of the MRS. Even if the field distortion is not negligible, we implicitly incorporate this distortion term in the \al distortion model. The absolute \al coordinate does not need to be correct in this step since we are mainly interested in $\Delta \alpha$, the distance of the points along the \al coordinate in order to calibrate the slice specific distortion. The final \al/\be system is defined after the new \al distortion is derived.

\subsection{Along-slice \al distortion calibration}

With the relative astrometry derived from the V2/V3 transform derived in section \ref{sec:flightFoV}, we can measure and fit the distortion in the iso-\al traces for every slice in each of the 12 spectral sub-bands (1A-4C). As described in Section~\ref{sec:det_fit}, with the forward modeling of the PSF we are able to fit the signal on the detector even for slices that do not contain the core of the PSF. We fit the \al distortion with a (Nx, Ny) = (2, 4) order polynomial. These values for the polynomial provided enough flexibility to fit the distortion everywhere on the detector without suffering from over-fitting especially at the edges of slices. Finally, we set the local MRS coordinate field to be centred around zero. We calculate the \al FoV limits for each slice using the detector slice mask and the newly derived \al distortion, and subtract the mean FoV value from the intercept coefficient of the polynomial transform of each slice.

\subsection{Local MRS to JWST V2/V3 coordinates transform}

\begin{figure}
    \centering
    \includegraphics[width=0.49\textwidth]{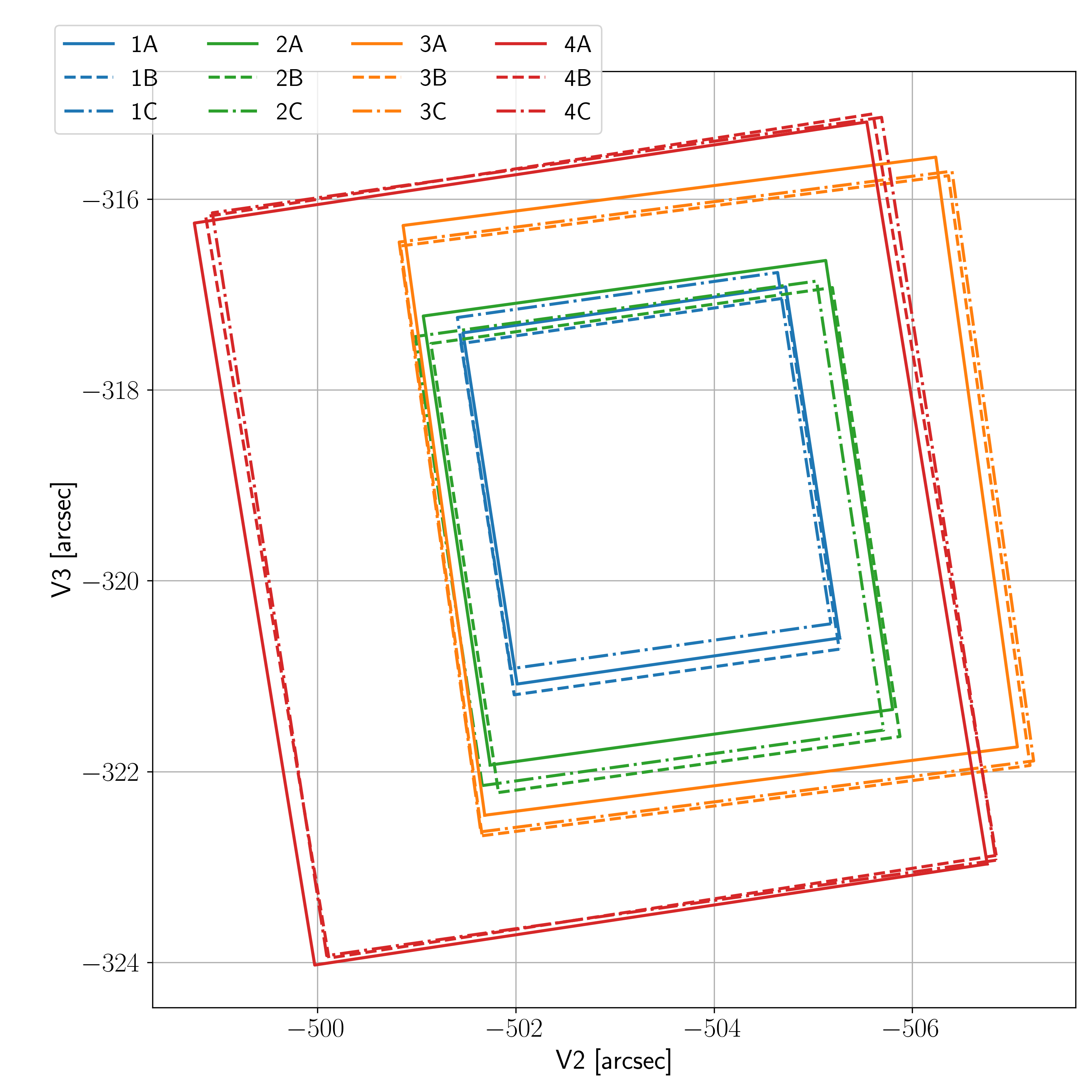}
    \caption{MRS band footprints in V2/V3 coordinates. This corresponds to the FoV of the MRS on sky, information that is used for pointing and dithering with the MRS.}
    \label{fig:MRS footprintsl}
\end{figure}

The last step of the analysis is to refine the V2/V3 to \al/\be transform, this time using the calibrated \al distortion. This yields important parameters for the boresight offsets of the MRS bands in the V2/V3 frame that is used by FGS to guide targets to the MRS and perform dithering offsets. The optimised dither pattern can be re-calculated given the new distortion and V2/V3 to \al/\be transform, and a polynomial transform shown in Eqs.~\ref{eq:v2_trans}, \ref{eq:v3_trans} is fitted. In Fig.~\ref{fig:MRS footprintsl} we show the final footprints of the MRS FoV of each band in the V2/V3 coordinate system, and in Table~\ref{tab:v2v3FoV} we list the relevant parameters for each band.

\begin{table}[h]
    \centering
    \caption{MRS field of view in global JWST coordinate system after finalising the distortion solution based on PID 1524. The angle refers to the rotation of the MRS slicer coordinate frame with respect to the JWST V3 coordinate.}
    \begin{tabular}{c|c|c|c|c}
    \toprule
        Band & V2 boresight & V3 boresight& angle & FoV (\al, \be) \\ 
        & [arcsec]& [arcsec]& [degrees] & [arcsec$^2$] \\ \hline
        1A & -503.37 & -319.00 & 8.42 & 3.3 x 3.7 \\
        1B & -503.36 & -319.12 & 8.30 & 3.3 x 3.7 \\
        1C & -503.30 & -318.85 & 8.30 & 3.3 x 3.7 \\
        2A & -503.43 & -319.29 & 8.16 & 4.1 x 4.8 \\
        2B & -503.51 & -319.57 & 8.24 & 4.1 x 4.8 \\
        2C & -503.35 & -319.50 & 8.17 & 4.1 x 4.8 \\
        3A & -503.96 & -319.01 & 7.59 & 5.4 x 6.2 \\
        3B & -504.01 & -319.21 & 7.59 & 5.6 x 6.2 \\
        3C & -504.02 & -319.17 & 7.59 & 5.6 x 6.2 \\
        4A & -502.75 & -319.61 & 8.73 & 6.9 x 7.9\\
        4B & -502.86 & -319.54 & 9.09 & 6.8 x 7.9\\
        4C & -502.89 & -319.54 & 8.43 & 6.8 x 7.9\\
        \bottomrule
    \end{tabular}
    
    \label{tab:v2v3FoV}
\end{table}

\subsection{Verification and estimated precision}

\begin{figure*}[h]
    \centering
    \includegraphics[width=0.85\textwidth]{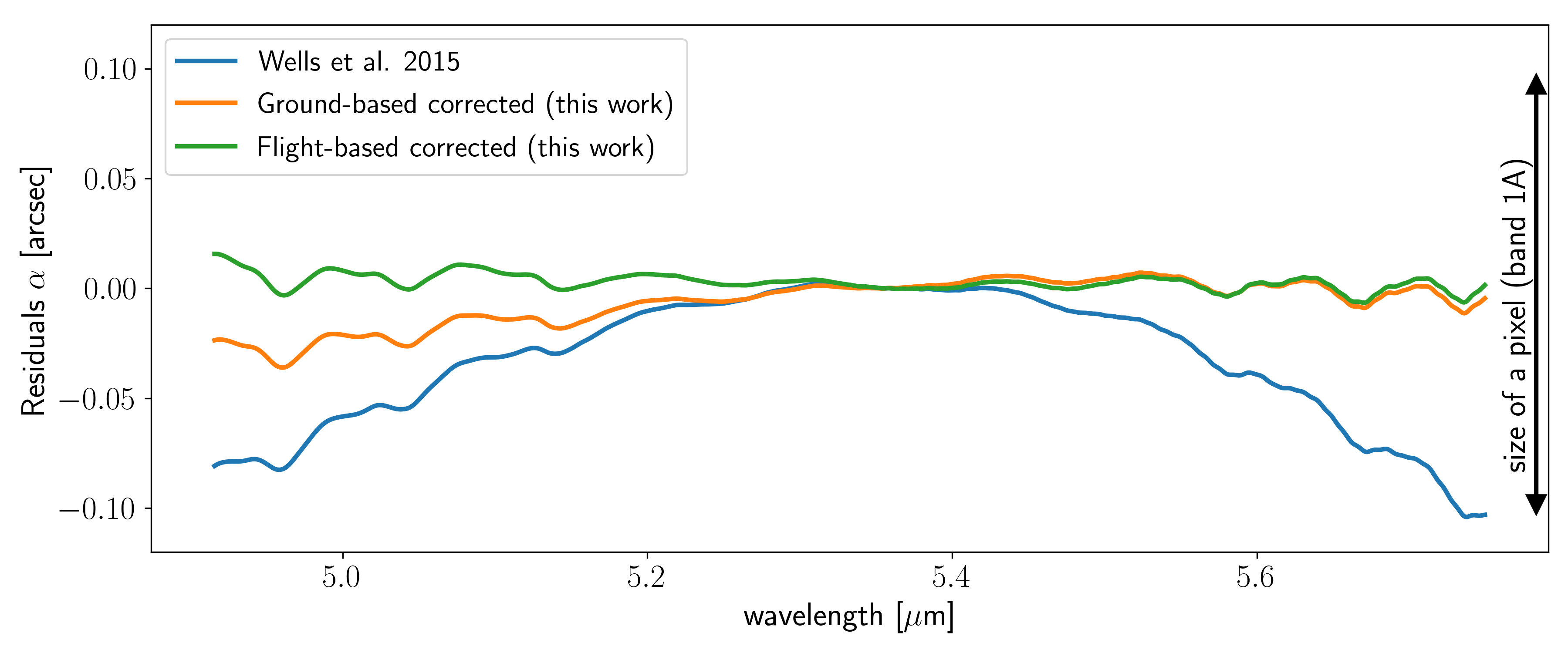} \\
    \includegraphics[width=0.85\textwidth]{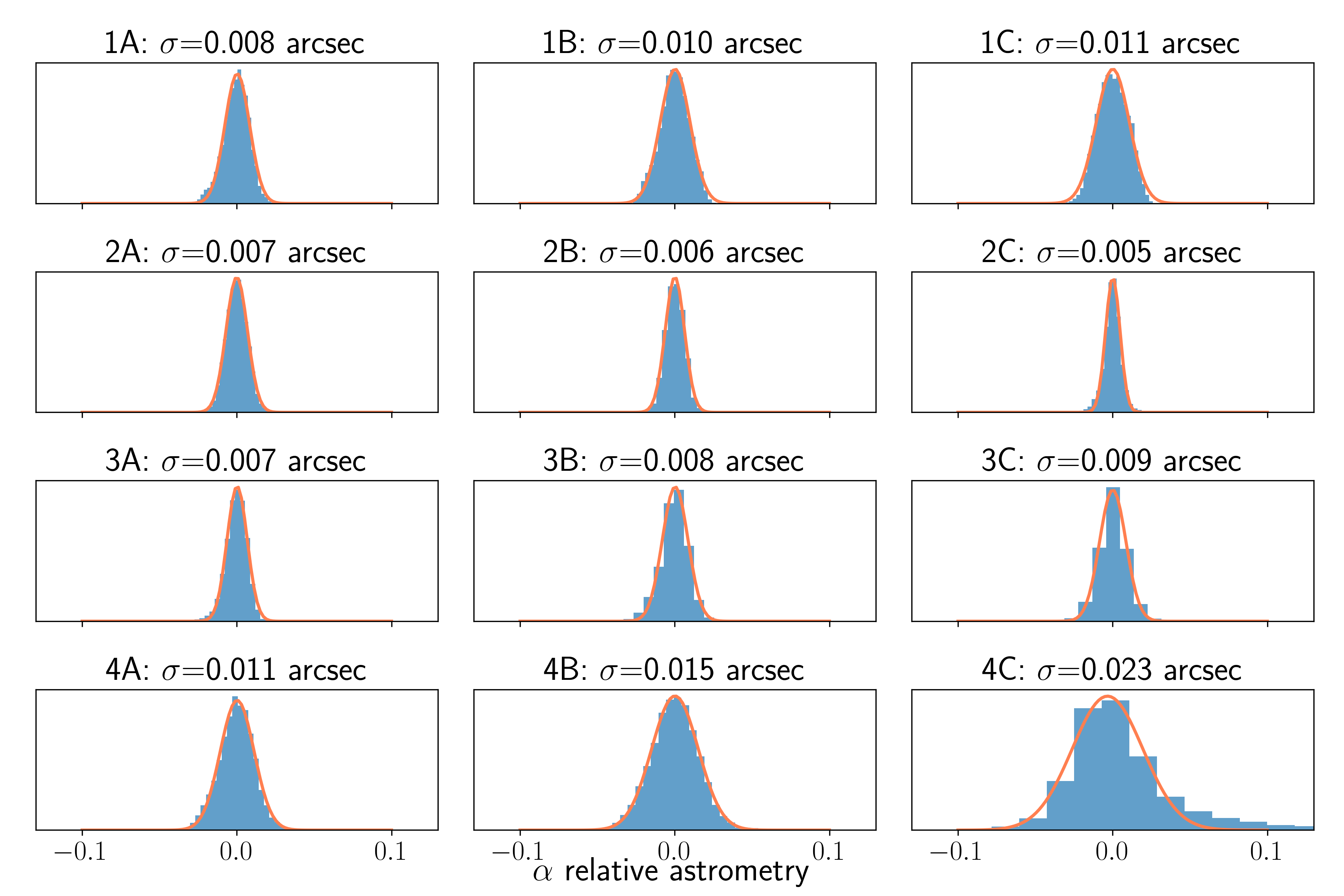}
    \caption{Relative distortion residuals. Top: a point source trace on the detector was extracted from PID 1050, and the distortion solution from \cite{Wells2015}, and this work was used to transform the iso-\al trace into \al, \lam coordinates. Since the point source should have a fixed spatial coordinate, this illustrates the improvement in the distortion model using the flight data. The residual modulation originates from the under-sampling of the MRS PSF. Bottom: distribution of iso-\al residuals for all dither point of PID 1049 and PID 1050, obtained by applying the distortion transforms derived in this work.}
    \label{fig:distortion_error}
\end{figure*}

To test the validity of the full chain of calibration we use other point source observations from PID 1049 and PID 1050, comparing their predicted locations in \al/\be to the ones measured from the data. The results are shown in Fig.~\ref{fig:distortion_error}. On the top panel, a single observation of the A-star from PID 1050 was used to compare the iso-\al trace. For each row the centre of the point source is fitted using the voigt profile and transformed into \al, \lam using the distortion solution from \cite{Wells2015}, the pre-launch distortion (Appendix~\ref{ap:ground_calibration}), and the distortion derived from flight data. Since the star should have a fixed spatial coordinate (\al, \be) as a function of wavelength, this illustrates the improvement of the latest distortion model. The residual modulation of the iso-\al trace that is seenIn the bottom panel of Fig.~\ref{fig:distortion_error},  we plotted the distribution of the residual between expected \al based on the pointing telemetry, and the measured iso-\al trace from the detector for all (10-13) individual dithers per band. Since we are interested in the relative error within a given MRS band we subtract the mean residual, such that the distributions are centred around zero. This removes any global offset which can arise from pointing errors or repeatability issues discussed in Section~\ref{sec:wheel_rep}. The standard deviation of a fitted Gaussian distribution to the residuals is reported, satisfying in all bands the tenth of a resolution element goal set out for the MRS \al distortion.

\begin{figure}[h]
    \centering
    \includegraphics[width=0.5\textwidth]{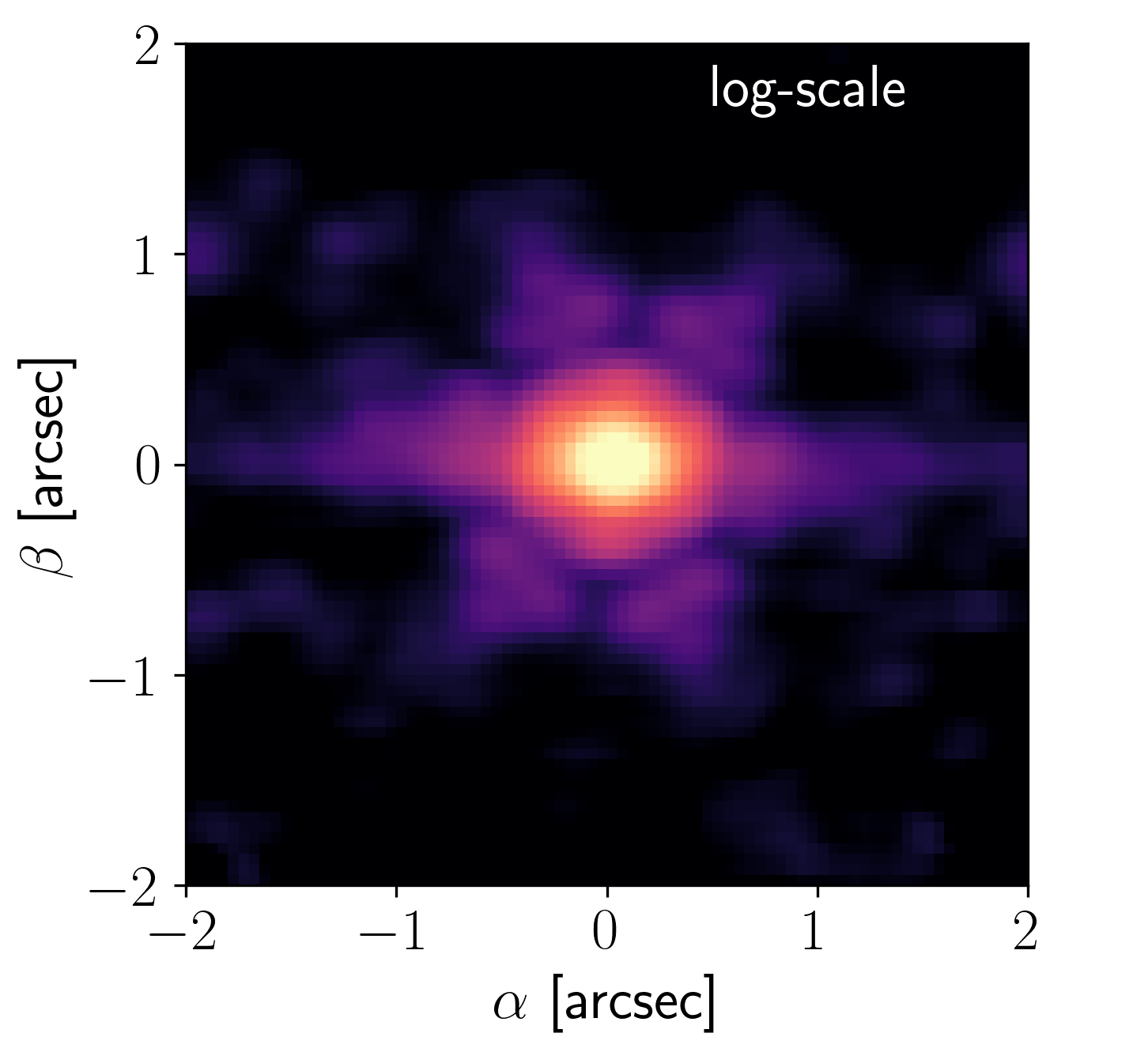}
    \caption{MRS Optical quality (band 1A) with data from PID 1050 enabled by the distortion solution derived in this work. The cube building step \citep{Law2023} uses a spaxel size of 0.05" and a wavelength bin of 0.05 \mum with the drizzle algorithm in IFUALIGN mode, that builds the cube in the \al, \be coordinate frame. The diffraction pattern (JWST ``petals'') of the PSF is revealed.}
    \label{fig:opt_q}
\end{figure}

Another example of illustrating that the distortion solution works as intended is to visualise a point source in the reconstructed cube. In Fig.~\ref{fig:opt_q} we show an A star (PID 1050, observation 9) in band 1A, with all 10 dithers positions combined with the size of the pixel in the cube (also referred to as spaxel, see \cite{Law2023} for details) set to 0.05" and binned over a broad wavelength bin of 0.05 \mum in order to increase the signal to noise ratio. This oversampling of the MRS PSF is possible because of the number of dithers available in the given observation, but also the fact that a broad wavelength bin naturally samples the PSF better due to the iso-\al trace being curved on the detector. The JWST PSF is revealed in detail, and we also observe the wings of the PSF at high S/N.

\section{Discussion} \label{sec:discussion}

\subsection{Repeatability of the Grating Wheel Assembly}\label{sec:wheel_rep}

During commissioning it was found that the repeatability of the angular positioning of the wheel induced measurable astrometric offsets. Each time the wheel moves between observations to select a different dichroic filter and grating setting, it returns to a given position with some uncertainty of around 0.5-1 pixels projected on the detector, based on the MIRI optical model. For example, the NIRSpec grating wheel also shows similar repeatability performance, which was already known from ground testing \citep{Boeker2023} and mostly affects the dispersion direction. For NIRSpec, two sensors measure this offsets and accordingly adjust the distortion model to compensate for it \citep{deOliveira2022_NIRSpec}. Unfortunately for the MRS, the positioning sensors do not have the required resolution to measure the changes in the wheel position.


In order to assess the repeatability of the astrometric solution we compared results across $\sim 10$ different observations of standard stars obtained throughout the Cycle 1 Calibration program using data from PIDs 1050, 1524, 1536, and 1538.  Each of these programs used target acquisition to observe well-known point sources in all twelve MRS bands, and all used a standard dither pattern (except PID 1524 Observation 16, which used the custom 57-point pattern). For each observation, we constructed a full 5-28 micron data cube and measured the centroid location of the target star as a function of wavelength in this cube.

\begin{figure*}[h]
    \centering
    \includegraphics[width=\textwidth]{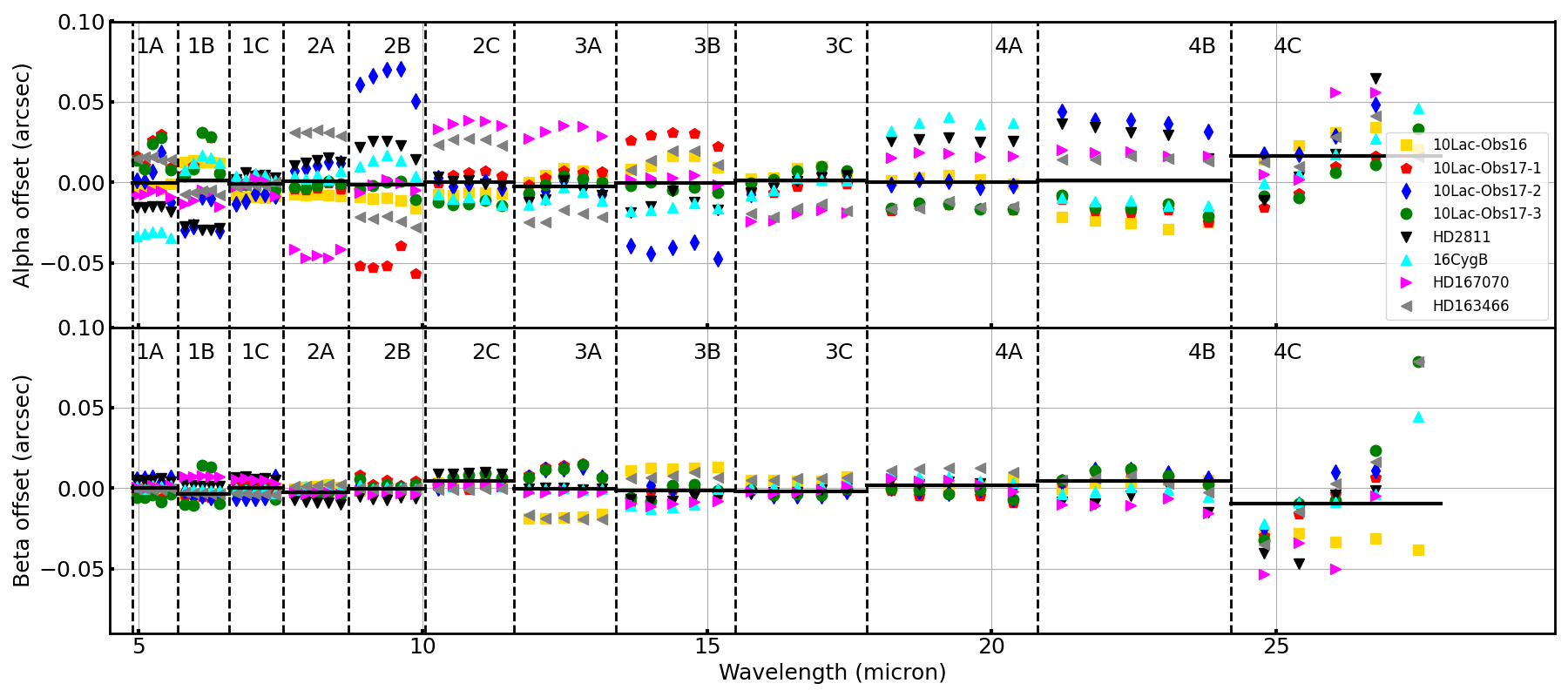}
    \caption{Change in the PSF centroid location as a function of wavelength for 5-28 $\mu$m data cubes constructed from dithered observations of a variety of standard stars observed by PIDs 1050, 1524, 1536, and 1538 (processed using the distortion solution derived in this work). The top panel shows offsets along the $\alpha$ direction, and the lower panel offsets along the $\beta$ direction.  Three different sets of data points are given for 10 Lac Observation 17 as these data were obtained during three complete rotations of the DGA assembly while the guide star remained locked explicitly to test the position repeatability of the DGA.  While the typical rms is much better than 10 mas for the $\beta$ direction, repeatability in the $\alpha$ direction is about 30 mas rms.}
    \label{fig:wheel_rep_astrometry}
\end{figure*}

We plot the result in Figure \ref{fig:wheel_rep_astrometry}, showing the relative offset from the median position in the alpha and beta directions as a function of wavelength.  While the $\beta$ direction positions are extremely stable from band to band, we note more significant jumps between individual bands in the $\alpha$ direction, with an rms of about 30 milliarcseconds.  This represents roughly 1/5 detector pixel or better repeatability overall, albeit with some extreme cases in which the offsets can be as large as about 1/2 pixel.  In Channel 2B for instance, we note a significant difference between -0.05 arcsec and +0.07 arcsec between successive observations of 10~Lac (blue and red points) during which the observatory remained in fine guide status and the only mechanism that moved between two sets of dithers was the DGA wheel. Another fact pointing towards the wheel repeatability being the issue is the fact that the Channel 1-4 and 2-3 offsets are anti-correlated, as one would expect from the optics of the MRS. The gratings of these pairs (1-4, 2-3) are respectively located on the same DGA wheel, and due to an additional reflection for channels 3 and 4, an offset in angle of the wheel will move the beam towards the opposite direction compared to channels 1 and 2.

For calibration purposes, we fit the average offset from the mean for each of the twelve bands and applied this boresight offset to the latest distortion reference files. This ensures that the astrometric location of a star is on average consistent across all MRS bands, even if repeatability can sometimes cause deviations from this mean position for individual observations.

The impact of the wheel repeatability on the data reduction and science products of the MRS was tested using the repeated measurement of PID 1524 (observation 17). We compared the exposures for band 2B which showed the worst offset  (Fig.~\ref{fig:wheel_rep_astrometry}, 2B blue and red points). No measurable change in the calibrated iso-\al trace, besides a constant astrometric offset, is found. Presumably, the displacement of the beam due to the diffraction grating is still within the linear regime of the distortion. Therefore, with no significant distortion effects, there are some minor issues that arise from the non-repeatability of the source on the detector between different dichroic settings. First, for extended sources where the astrometry cannot be measured in the data, the alignment of the bands in the 3D cube will not be exact and spatial features will appear to shift between bands. Due to the wavelength overlap between the bands, such cases can potentially be mitigated by examining the common wavelengths and the behaviour of any features from one band to the other. Second, dedicated point source fringe flats based on calibration observations would not be directly applicable since the point source fringes change rapidly from pixel to pixel on the detector \citep{Argyriou2020a,Gasman2022}. Third, astrometric measurements of the same source from different epochs might be biased and should include the repeatability error in its uncertainty. Finally, due to the relatively small FoV, complex scenes involving extended structures of interest or multiple sources need to be planned more carefully, taking into account the additional 100 mas astrometric uncertainty.

\subsection{Lessons learned for future calibration}

Due to its design and operating wavelengths, the MRS is a challenging instrument to calibrate. A series of instrumental effects are present due to a combination of design choices and peculiarities of the MIRI detectors, such as spectral fringing, scattered light within the detector, under-sampling of the PSF, and optical distortion \citep{Argyriou2023_MRS}. The following list identifies some of the insights for testing an IFU, after a decade of calibration work performed by the instrument team.

\begin{itemize}
    \item Testing the MRS in the lab, at conditions representative of its flight status is challenging. The MIRI Telescope Simulator, built specifically for testing MIRI, could not provide a bright diffraction limited point source at all wavelengths, due to a combination of optical distortions of the test setup and lack of flux of the test light source. Therefore, the early investment and development of a robust testing facility is key. 
    
    \item For calibrating the distortion and astrometry pre-flight, a well defined point source that can be steered precisely in the FoV of the IFU is required, correlated to a distortion free reference frame. This could potentially be the steering stage X, Y positions. Even with its limitations the MTS provided many insights into the distortion of the MRS, and was able to deliver a preliminary calibration.
    
    \item Hardware tested in flight will potentially reveal new instrumental systematic errors that were not caught in the instrument testing on ground. Therefore, planning for dedicated commissioning observations for all key parameters is essential. Additionally, allowing for flexible project management of commissioning was very valuable in order to adapt observations and tests given the first data that were obtained. 
    
    \item Finally, careful planning and development of the optimised dither pattern used for PID 1524 (observation 16) enabled the calibration of the  distortion efficiently using a bright star. Additionally the same dataset could be used for multiple additional tests like wavelength, fringing, optical quality, and photometric calibration as a function of the position in the FoV.
    
\end{itemize}

\subsection{Summary} 

We have derived the full astrometric distortion solution of the MRS, based on flight data. The calibration of JWST field coordinates V2/V3 to the local MRS slicer coordinates \al/\be is required for the operation of the MRS, used during pointing and dithering of observations. The transform from detector to \al/\be and subsequently V2/V3, enables the pipeline to correct for the optical distortion introduced due to the IFUs, a required step for the rectification of the 2D detector data into 3D cubes, that reconstruct the observed astronomical scene and allow for the extraction of spectra. A good understanding of the distortion might enable modelling of the MRS PSF and application of optimised forward modelling approaches for MRS data. The final precision of the distortion solution ranges from 10 mas in channels 1-3, up to 23 mas in channel 4. The total astrometric uncertainty of a given MRS exposure with target acquisition is estimated to be 50 mas, given by $\sqrt{\sigma_{distortion}^2+\sigma_{TA}^2+\sigma_{repeatability}^2}$.

\begin{acknowledgements}
\newline
\newline
   \textbf{Contributions:} PP lead the calibration work package for the MIRI MRS distortion and astrometry, performed the main analysis, and wrote the manuscript. IA co-lead the calibration and analysis of the distortion, and derived the MRS slice mask. DL planned the dedicated cycle 1 calibration program 1524, performed the astrometric monitoring of the MRS, and implemented the calibation products in the JWST pipeline. IA, DL contributed to the manuscript writing and editing. DL, AMG, AG, AL contributed to the overall calibration and analysis of the distortion. All authors read and provided feedback on the manuscript.

\newline
\newline

This research has made use of the NASA Astrophysics Data System and the \texttt{python} packages \texttt{numpy} \citep{harris2020array}, \texttt{scipy} \citep{2020SciPy-NMeth}, \texttt{matplotlib} \citep{Hunter:2007} and \texttt{astropy} \citep{astropy:2013, astropy:2018}.

\newline
\newline
   The authors want to thank the hundreds of engineers and scientists, whose contributions over 25 years made the JWST mission possible.
   Polychronis Patapis thanks the Swiss Society for Astrophysics and Astronomy (SSAA) for the MERAC Funding and Travel award.
   Ioannis Argyriou, Danny Gasman, and Bart Vandenbussche would like to thank the European Space Agency (ESA) and the Belgian Federal Science Policy Office (BELSPO) for their support in the framework of the PRODEX Programme. 
   Patrick J. Kavanagh acknowledges support from the Science Foundation Ireland/Irish Research Council Pathway programme under Grant Number 21/PATH-S/9360. 
   Javier~Álvarez-Márquez acknowledges support by grant PIB2021-127718NB-100 by the Spanish Ministry of Science and Innovation/State Agency of Research MCIN/AEI/10.13039/501100011033 and by “ERDF A way of making Europe”
   Alvaro Labiano acknowledges the support from Comunidad de Madrid through the Atracción de Talento Investigador Grant 2017-T1/TIC-5213, and PID2019-106280GB-I00 (MCIU/AEI/FEDER,UE). 
   The work presented is the effort of the entire MIRI team and the enthusiasm within the MIRI partnership is a significant factor in its success. MIRI draws on the scientific and technical expertise of the following organisations: Ames Research Center, USA; Airbus Defence and Space, UK; CEA-Irfu, Saclay, France; Centre Spatial de Liége, Belgium; Consejo Superior de Investigaciones Científicas, Spain; Carl Zeiss Optronics, Germany; Chalmers University of Technology, Sweden; Danish Space Research Institute, Denmark; Dublin Institute for Advanced Studies, Ireland; European Space Agency, Netherlands; ETCA, Belgium; ETH Zurich, Switzerland; Goddard Space Flight Center, USA; Institute d'Astrophysique Spatiale, France; Instituto Nacional de Técnica Aeroespacial, Spain; Institute for Astronomy, Edinburgh, UK; Jet Propulsion Laboratory, USA; Laboratoire d'Astrophysique de Marseille (LAM), France; Leiden University, Netherlands; Lockheed Advanced Technology Center (USA); NOVA Opt-IR group at Dwingeloo, Netherlands; Northrop Grumman, USA; Max-Planck Institut für Astronomie (MPIA), Heidelberg, Germany; Laboratoire d’Etudes Spatiales et d'Instrumentation en Astrophysique (LESIA), France; Paul Scherrer Institut, Switzerland; Raytheon Vision Systems, USA; RUAG Aerospace, Switzerland; Rutherford Appleton Laboratory (RAL Space), UK; Space Telescope Science Institute, USA; Toegepast- Natuurwetenschappelijk Onderzoek (TNO-TPD), Netherlands; UK Astronomy Technology Centre, UK; University College London, UK; University of Amsterdam, Netherlands; University of Arizona, USA; University of Bern, Switzerland; University of Cardiff, UK; University of Cologne, Germany; University of Ghent; University of Groningen, Netherlands; University of Leicester, UK; University of Leuven, Belgium; University of Stockholm, Sweden; Utah State University, USA. A portion of this work was carried out at the Jet Propulsion Laboratory, California Institute of Technology, under a contract with the National Aeronautics and Space Administration.

   We would like to thank the following National and International Funding Agencies for their support of the MIRI development: NASA; ESA; Belgian Science Policy Office; Centre Nationale D'Etudes Spatiales (CNES); Danish National Space Centre; Deutsches Zentrum fur Luft-und Raumfahrt (DLR); Enterprise Ireland; Ministerio De Economiá y Competividad; Netherlands Research School for Astronomy (NOVA); Netherlands Organisation for Scientific Research (NWO); Science and Technology Facilities Council; Swiss Space Office; Swedish National Space Board; UK Space Agency.

   We take this opportunity to thank the ESA \textit{JWST} Project team and the NASA Goddard ISIM team for their capable technical support in the development of MIRI, its delivery and successful integration.

\end{acknowledgements}

\bibliography{refsastrometry} 
\bibliographystyle{aa} 

\begin{appendix}

\section{Pre-launch calibration}\label{ap:ground_calibration}

\subsection{Ground test data}
\textbf{MIRI Flight Model Test Campaign.}

The MIRI Flight Model (FM) test campaign took place at the Rutherford Appleton Laboratory in the UK in 2011. The MIRI instrument and its operational modes were tested under cryogenic conditions, and dedicated test runs were performed to characterise all aspects of the instrument (optical alignment, performance of electronics and detectors, optical transmission, etc). The FM test campaign included a fine raster scan of a point source with multiple exposures per slice for each sub-band of the MRS, dithered observations for the FoV measurement, and observations with a spatially uniform extended source. A dedicated optical system, the MIRI Telescope Simulator \citep[MTS,][]{mts_paper, Herrada2007}, was built in order to provide the necessary input illumination. A black body source was available operated at a temperature of 400~K, 600~K and 800~K. For simulating point sources, a 100 \mum pinhole mask was placed in the pupil wheel of the MTS and provided a point-like source that could be steered with high precision. 

It should be noted that due to a non-uniform illumination of the pupil at the field location of the MRS, the pinhole mask did not produce a diffraction limited PSF but rather an elongated semi-extended source. The flux from the source was also relatively low, with a maximum of 3 Digital Numbers per second (DN/s), out of the $\sim$2000 DN/s that an almost saturating source would provide. This resulted in low signal to noise ratio (S/N) in the raster and FoV observations of the point source. Finally, the most critical issue for the geometric distortion calibration based on FM data was a field distortion of the MTS itself that introduced a systematic error in the reference coordinate system of the MTS source (MTS-X/MTS-Y). This distortion was measured from MIRI Imager exposures and modeled in Zemax OpticStudio, indicating a significant shear in the MTS field which could not be corrected to the precision required for the MRS distortion calibration. Nonetheless, the FM data provide a starting point for deriving the intra-slice distortion, with a raster scan covering all slices and bands of the MRS. Even with a distorted MTS field, all the spectral bands can be placed onto a common coordinate frame.


\textbf{MIRI Cryogenic Vacuum Test Campaign}

Three Cryogenic Vacuum (CV) test campaigns were conducted at Goddard Space Flight Center between 2013 and 2016, denoted CV1, CV2 and CV3 \citep{ISIM2016}. The whole JWST Integrated Science Instrument Module (ISIM) was placed in a cryostat and cooled to operating temperature (this did not include the JWST telescope optics or bus). The testing setup provided a JWST-like point source with high flux but in a very limited wavelength range. For the MRS the useful signal was limited between 4.9--6.1~\mum. The dither patterns observed were associated with the ISIM coordinate system defined by coordinates XAN/YAN, which provided a robust and distortion-free reference frame. Besides a 9-point dither pattern for the PSF measurement in CV2 and CV3, a very useful across slice scan was performed in CV3, where the source was displaced in fixed steps over a slice. This last dataset is used to evaluate our fitting routines as described in Sect.~\ref{sec:det_fit}.

\textbf{Ground data processing}

The FM and CV data were taken from the Rutherford Appleton Laboratory (RAL) archive, already processed to slope images (units of Digital Numbers per second) \citep{morrison23}. All FM and CV data had dedicated backgrounds associated with the science exposures, which were always subtracted.

\subsection{Astrometric reference frame using sub-band 1A}\label{sec:ref_frame1A}

At the beginning of the analysis to derive the MRS geometric distortion, we address the fact that there is a lack of a reference coordinate system that can connect the FoV of all 12 MRS bands and provide the distance of the FM raster scan points in \al/\be coordinates. These distances would then be fitted to estimate the polynomial coefficients of the distortion transform Eq.~\ref{eq:alpha_polynomial}. Ideally, this reference frame would have been the MTS commanded coordinates (MTS-X/MTS-Y), but due to the distortion of the MTS field itself, the reference would have been biased. First, the MTS distortion needs to be taken into account, and therefore, we consider the XAN/YAN coordinate frame of the CV2 FoV measurement. The XAN/YAN coordinates are assumed to be non-distorted and the CV2 measurement provided 9 points in the FoV of 1A. Unfortunately band 1A is the only band with enough signal from the ISIM point source. 

The dither pattern of CV2 was based on the distortion derived from the FM data, and therefore the commanded offsets could not a priori be assumed to be orthogonal to the slicer of 1A.  We used the sampling of CV2 point source by the slices to determine the centre of each pointing in the across-slice position, in units of slice number $s$. We then performed an affine transform to offset, rotate and scale the XAN/YAN coordinates to \al/\be. The affine transform is given by Eq.~\ref{eq:field_transform2} where Eq.~\ref{eq:field_transform2} is least-squares minimized in order to estimate the parameters $\beta_0$, $\Delta \beta$, $\theta$, $X_0$ and $Y_0$.



The parameters $\beta_0$ and $\Delta \beta$ fully define the \be coordinate of the local MRS slicer coordinate system, and \al is just defined orthogonal to that. The angle $\theta$ provides the rotation of the MRS 1A FoV with respect to the JWST coordinate system V2/V3. For the points of the CV2 campaign, we therefore have a valid and unbiased reference frame. 

Next, we made use of the Zemax OpticStudio optical model to obtain a set of calibrated slices in the locations where the CV2 data had available points. Since we effectively only have two points per slice from the CV2 FoV measurement, we can only correct the \al distortion up to a magnification but cannot check or correct the plate scale gradient (the second order term). By examining the Zemax OpticStudio model prediction as it is, we see that the projected slices on the detector are offset compared to the slice mask derived in Sect.~\ref{sec:slice_mask}. We correct that by moving the Zemax OpticStudio slices on the middle of the detector (row 512) to match the measured slice edges. This yields a $\Delta x$ value for each of the slices in band 1A, corresponding to the shift in detector x between Zemax OpticStudio and measured slices. 

\begin{figure}
    \centering
    \includegraphics[scale=0.35]{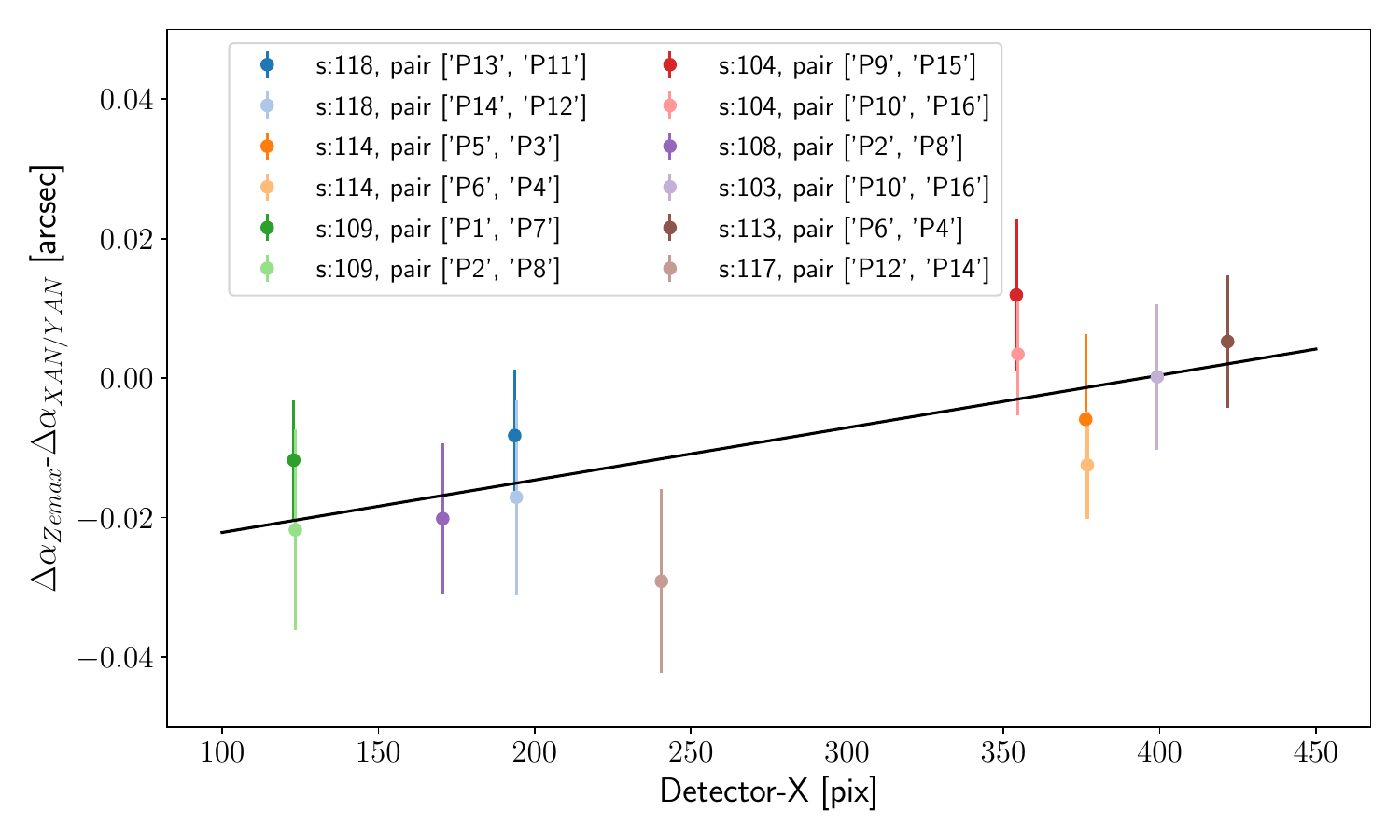}
    \caption{Zemax alpha distortion calibration}
    \label{fig:zemax_sys}
\end{figure}

For each of the slices that have at least two point sources we identified pairs of points that have approximately the same \be coordinate, and used their \al coordinates from the XAN/YAN transform to calculate their distance, $\Delta \alpha_{XAN/YAN}$,  in units of arcsec. We then measured their iso-\al on the detector, using the Zemax OpticStudio model we obtained \alz, and again calculated the distance, denoted $\Delta \alpha_{Zemax}$. In Fig.~\ref{fig:zemax_sys} we show the difference between $\Delta \alpha_{XAN/YAN}$ and $\Delta \alpha_{Zemax}$ as a function of position along detector columns. This difference reflects a discrepancy in the magnification of the field in the Zemax OpticStudio model, which we correct by multiplying \alz by a scaling factor for each slice. Interestingly there seems to be a correlation between detector-X position of the slice and the magnification error. Another systematic effect that can be seen in Fig.~\ref{fig:zemax_sys} is a difference between pairs in the same slice. The darker colours of the points refer to the point source pairs that are well centred in the slice, while the lighter colors correspond to the half-slice offset of the dither pattern, with the point source centred in between two slices. Finally, there is a slice dependent offset between Zemax OpticStudio and the reference frame which is expected from slight slit mask offsets which would cover different parts of the field for each slice. This is corrected with a constant \al offset for each of the slices, concluding the calibration of slices 3, 4, 8, 9, 13, 14, 17 and 18 (21 slices in total).

\subsection{MTS to reference frame transform}

\begin{figure}
    \centering
    \includegraphics[scale=0.5]{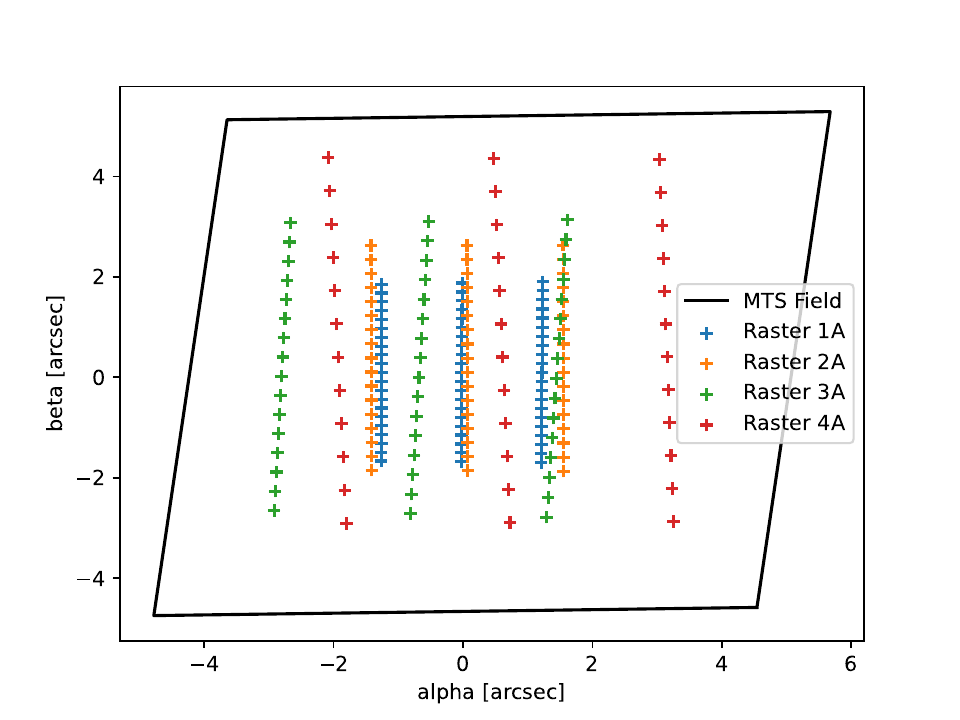}
    \caption{MTS field projected onto \al/\be of band 1A. The raster scan point of bands 1A/2A/3A/4A are plotted, which are used for measuring the distortion of the MRS.}
    \label{fig:MTS_to_albe}
\end{figure}

With several calibrated slices in band 1A, but lacking any reference frame in other bands, we proceeded to derive a transform linking the MTS coordinates to the reference frame defined in Sect.~\ref{sec:ref_frame1A}. This step is essential for enabling the distortion calibration of all other bands (that did not have any useful signal in the CV campaign). We used the main raster scan exposures from the FM campaign, selected the calibrated slices and measured their \al/\be coordinates from the detector. We then fitted the affine transform in Eq.~\ref{eq:mts_field_transform}, including translation, rotation, scale and shear, to map the MTS-X/MTS-Y coordinates to \al and \be respectively. In Fig.~\ref{fig:MTS_to_albe} we show a square field in MTS coordinates projected onto reference 1A \al/\be frame, as well as the raster scan pointings of each of the SHORT bands 1A, 2A, 3A, 4A. 

\begin{equation} \label{eq:mts_field_transform}
    \begin{bmatrix}
    \alpha \\
    \beta
    \end{bmatrix}  
    =\begin{bmatrix}
     \alpha \\
     \beta_0 + (s-1)*\Delta \beta
    \end{bmatrix}\\
    = R[\theta]S[\textbf{s}]
    \begin{bmatrix}
    MTS_X-X_0\\
    MTS_Y-Y_0
    \end{bmatrix},
\end{equation}

where 
\begin{equation*}
    R[\theta] = \begin{bmatrix}
     cos(\theta) & sin(\theta) \\
     -sin(\theta) & cos(\theta)
    \end{bmatrix} \,\text{and}\,
    S[\textbf{s}] = \begin{bmatrix}
     \sigma_x & s_x \\
    s_y & \sigma_y
    \end{bmatrix}.
\end{equation*}
The parameters $\sigma_{x/y}, s_{x/y}$ correspond to a field magnification and shear between MTS and the local MRS coordiantes.

With the raster scans mostly aligned to the IFU image slicer for each band, we can observe the slight differences between the band-to-band boresight and rotation of the slicers. Using a similar model as in Eq.~\ref{eq:field_transform1}, we calculate the transform from the reference frame 1A to the local \al/\be frame for each band, based on the detector slice coordinates of the raster scans. This yields the $\beta_0$ and $\Delta \beta$ parameters for defining the \be coordinate, as well as the relative rotation and \be boresight offset with respect to band 1A, for all 12 sub-bands. The final value of the boresight shift in \al between the bands were determined after the final calibration of the alpha distortion in Sect.~\ref{sec:flightFoV}. 


\subsection{Along-Slice Distortion Transformation Matrix}\label{sec:d2c_trans}

The procedure was the same for each band. The main raster scan of the given band was used, with each slice having available three exposures centred on the slice. The iso-\al trace on the detector was fitted for each exposure using a Gaussian profile on the detector as detailed in Sect.~\ref{sec:det_fit}, going up the detector rows. The derived MTS to local MRS transforms provide the required astrometric information to assign each of the raster scan exposures an \al coordinate based on their MTS coordinates. The detector fringes modulate the \al centre along the dispersion direction and affect its estimation, and since we expect the distortion to be a smooth function we fitted a two-dimensional polynomial of order 2 in detector-X and order 4 in detector-Y. 

\subsection{Iteration of distortion solution}

\begin{figure}[h]
    \centering
    \includegraphics[width=\columnwidth]{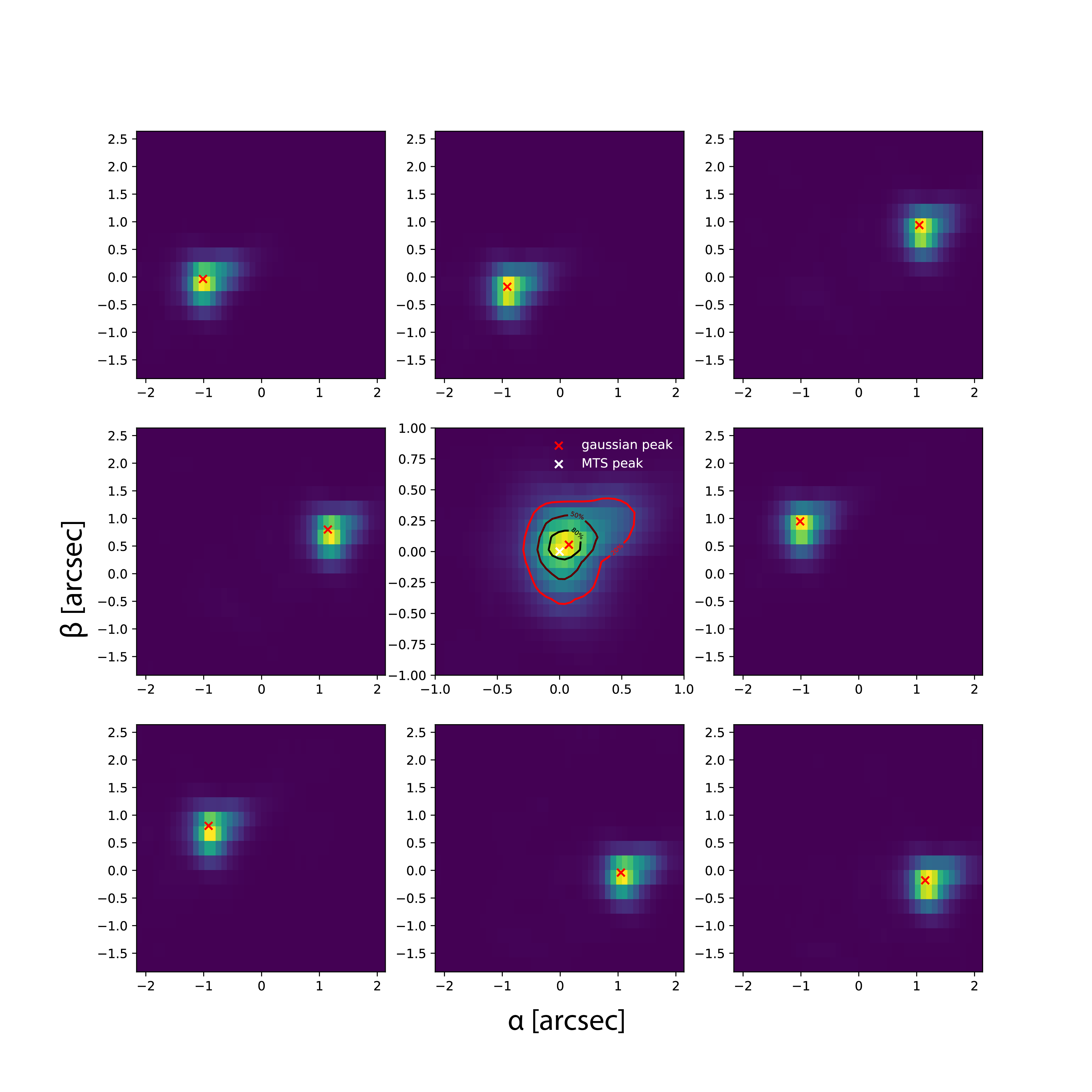}
    \caption{Estimation of the MTS PSF from the FM PSF observations. The individual dithers are combined into a higher resolution PSF, revealing the shape of the MTS point source. This PSF can be used when iterating the distortion solution in order to improve centroid estimation.}
    \label{fig:FM_PSF}
\end{figure}

Having derived a first distortion solution based on the FM data, we used the FM PSF measurement to reconstruct the PSF in \al/\be space. To do so we need to interpolate the detector pixels onto the MRS local coordinate frame. We used an inverse distance weighting algorithm, with exponential decaying weights (referred to in the pipeline cube building as \texttt{emsm}\footnote{\url{https://jwst-pipeline.readthedocs.io/en/latest/jwst/cube_build/main.html}}). The offset of each dither was estimated using its MTS coordinates and transforming them into \al/\be coordinates. In Fig.~\ref{fig:FM_PSF} the reconstructed and over-sampled PSF of band 2A is plotted, together with the individually reconstructed dithered exposures. We observe that the PSF is semi-extended and does not show the expected diffraction limited pattern. Not only that, but for the purpose of deriving the distortion solution, as seen in Fig.~\ref{fig:FM_PSF}, the centroid estimated by a Gaussian does not match the centre of the PSF.

\begin{figure}[h]
    \centering
    \includegraphics[width=\columnwidth]{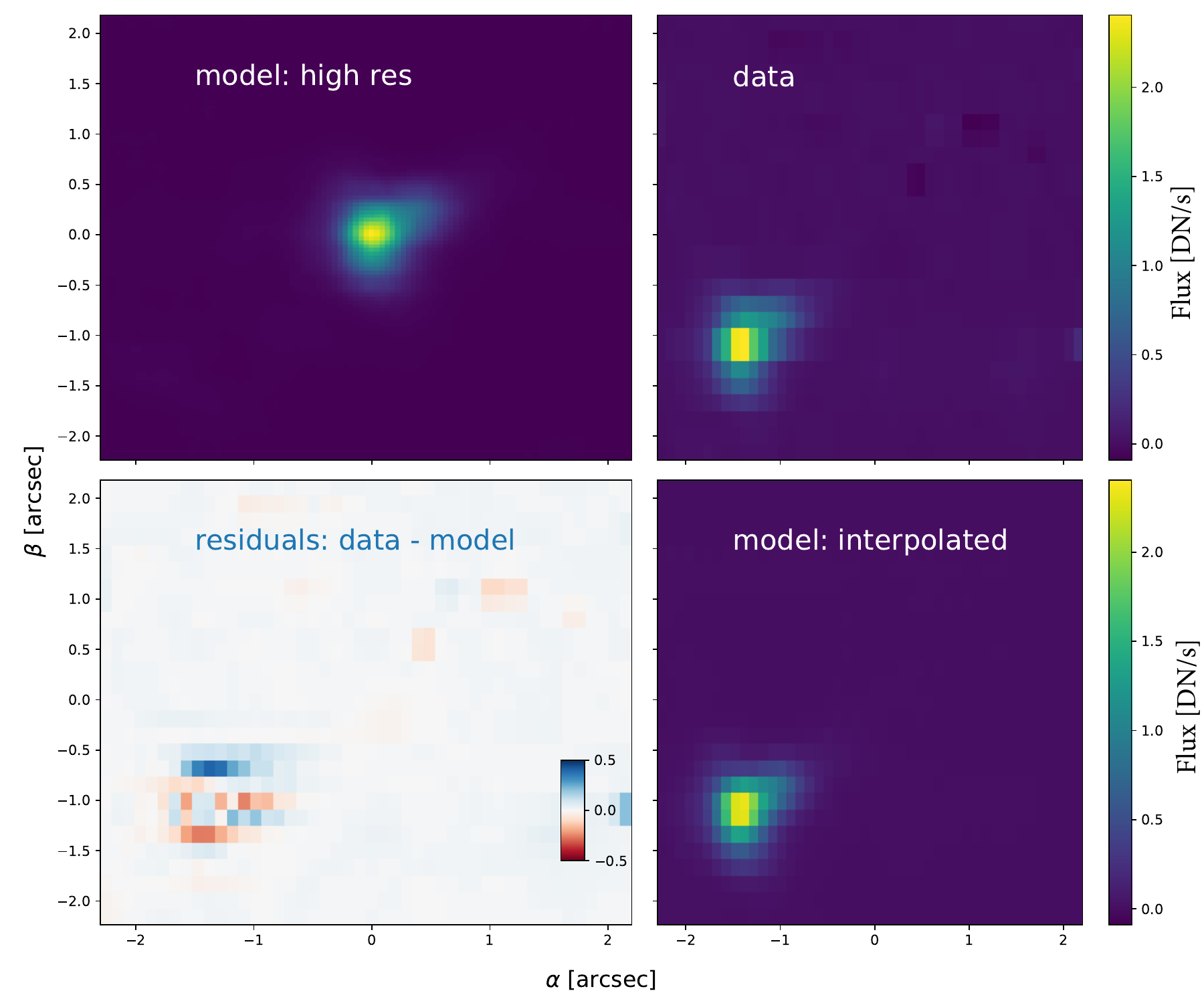}
    \caption{Forward modeling of the FM PSF used for the estimation of \al, \be centroids and the iso-\al trace. Top left: the high resolution interpolated model of FM PSF. Top right: a reconstructed cube of a single point source exposure using the initial distortion solution. Bottom right: model PSF interpolated on the grid of the cube to match the observed data. Bottom left: difference between data and projected PSF model.}
    \label{fig:FM_projection}
\end{figure}

Therefore, we iterate on the distortion solution by using the reconstructed PSF as a forward model to estimate the centre of a given FM point source observation. This activity was implemented in two ways, illustrated in Fig.~\ref{fig:FM_projection}. First, for band 1A, we used the 2D PSF to fit the points of the raster scan and to iterate on the MTS to \al/\be transform. The high resolution PSF is projected onto the local MRS coordinates with the \texttt{RegularGridInterpolator} from the \texttt{scipy} package using linear interpolation. The optimisation used the \texttt{L2 norm} as a loss function to minimise the difference between model and the data, with three free parameters of amplitude, offset \al, and offset \be. To ensure the minimisation converged, we restricted the evaluation of the loss function within an aperture of 0.5" around the point source ($\sim$3 slices on the image slicer). This process yielded new centroids which were in turn used to calculate the new transform with Eq.~\ref{eq:mts_field_transform}.

Second, using the distortion solution derived earlier, the PSF model is projected onto the detector for every row to estimate the iso-\al trace accounting for the shape of the PSF. This is repeated for every point in the raster scan, and the distortion was estimated again on the updated traces. The projection uses the estimate of the slice coordinate from the detector to provide the PSF \al-profile expected in the slice of interest. Typically this only works for slices either containing the peak of the PSF, or the slices adjacent to that, due to the signal to noise ratio of the FM data being very low. Here the \texttt{scipy} function \texttt{interp1d} was used, again with linear interpolation. 

With the iso-\al traces and MTS to \al/\be field transform based upon the FM PSF, we re-calculate all the distortion polynomial coefficients as in Sect.~\ref{sec:d2c_trans}.
\end{appendix}
\end{document}